\documentclass[conference]{IEEEtran}
 \usepackage{cite}
\usepackage{microtype}
\usepackage{graphicx}
\usepackage{booktabs} 
\usepackage{hyperref}

\usepackage{amsthm}

\usepackage[algo2e]{algorithm2e}
\usepackage{algorithm}
\usepackage{algorithmic}
\usepackage{color}
\usepackage{multirow}
\usepackage{dcolumn}
\newlength{\thinline}
\newlength{\thickline}
\usepackage{threeparttable}
\renewcommand{\figurename}{Figure}
\usepackage{url}
\usepackage{amsmath}
\newtheorem{theorem}{Theorem}
\newtheorem{lemma}{Lemma}
\usepackage{amssymb}

\usepackage{ulem}
\usepackage{subcaption}
\usepackage{xspace}
\newtheorem{myDef}{Definition}

\newcommand{\wqruan}[1]{\textcolor{black}{#1}}

\newcommand{\newchange}[1]{\textcolor{black}{#1}}
\newcommand{\share}[1]{\left \langle {#1} \right \rangle}

\floatname{algorithm}{Protocol}
\renewcommand{\algorithmicrequire}{\textbf{Input:}}
\renewcommand{\algorithmicensure}{\textbf{Output:}}
\newcommand{\queqiao}{\texttt{\texttt{Queqiao$_\epsilon$}}\xspace}
\newcommand{\tfencrypted}{\texttt{\texttt{TF-Encrypted$_\epsilon$}}\xspace}
\usepackage{algorithm}

\begin{document}

\title{\huge Private, Efficient, and Accurate: Protecting Models Trained by Multi-party Learning with Differential Privacy}

\author{Wenqiang Ruan$^\$$, Mingxin Xu$^\$$, Wenjing Fang$^+$,  Li Wang$^+$, Lei Wang$^+$, Weili Han$^\$$ \\\textit{$^\$$Laboratory of Data Analytics and Security, Fudan University, $^+$Ant Group}\\
\textit{$^\$$\{wqruan20, 20212010078, wlhan*\}@fudan.edu.cn}, 
\textit{$^+$\{bean.fwj, raymond.wangl, shensi.wl\}@antgroup.com
}
\\}

\maketitle


\begin{abstract}
Secure multi-party computation-based machine learning, referred to as multi-party learning (MPL for short), has become an important technology to utilize data from multiple parties with privacy preservation. While MPL provides rigorous security guarantees for the computation process, the models trained by MPL are still vulnerable to attacks that solely depend on access to the models. Differential privacy could help to defend against such attacks. However, the accuracy loss brought by differential privacy and the huge communication overhead of secure multi-party computation protocols make it highly challenging to balance the 3-way trade-off between privacy, efficiency, and accuracy.

In this paper, we are motivated to resolve the above issue by proposing a solution, referred to as PEA (Private, Efficient, Accurate), which consists of a secure differentially private stochastic gradient descent (DPSGD for short) protocol and two optimization methods. First, we propose a secure DPSGD protocol to enforce DPSGD, which is a popular differentially private machine learning algorithm, in secret sharing-based MPL frameworks. Second, to reduce the accuracy loss led by differential privacy noise and the huge communication overhead of MPL, we propose two optimization methods for the training process of MPL: (1) the data-independent feature extraction method, which aims to simplify the trained model structure;  (2) the local data-based global model initialization method, which aims to speed up the convergence of the model training. We implement PEA in two open-source MPL frameworks: \texttt{TF-Encrypted} and \texttt{Queqiao}. The experimental results on various datasets demonstrate the efficiency and effectiveness of PEA. E.g. when $\epsilon = 2$, we can train a differentially private classification model with an accuracy of 88\% for CIFAR-10 within 7 minutes under the LAN setting. This result significantly outperforms the one from \texttt{CryptGPU}, one state-of-the-art MPL framework: it costs more than 16 hours to train a non-private deep neural network model on CIFAR-10 with the same accuracy. 
\end{abstract}
  


\maketitle



\pagestyle{plain}
\setcounter{page}{0}
\setcounter{page}{1}
\section{Introduction}\label{intro}
Since European Union released General Data Protection Regulation~\cite{voigt2017eu}, data transfer has been strictly restricted more than before~\cite{9444564}. In order to utilize data stored in different parties with privacy preservation, secure multi-party computation (SMPC for short) based machine learning~\cite{DBLP:conf/sp/MohasselZ17,mohassel2018aby3,cryptGPU}, referred to as MPL~\cite{song2020sok}, is an important technology. Benefiting from the rigorous security guarantee provided by SMPC, through using MPL, data analysts can train machine learning models based on data from multiple sources without leaking any private information except for the trained models. 

However, many attacks~\cite{shokri2017membership, li2021membership}, e.g. membership inference attacks~\cite{shokri2017membership}, on the machine learning models themselves have been proposed in recent years. These attacks only require access to the trained models (i.e. no raw data, even no intermediate results). For example, by using membership inference attacks, attackers can infer the membership of one data point by querying the target model~\cite{shokri2017membership}. As these attacks are missed by the threat models of SMPC, they would seriously break the security of MPL frameworks.

Differential privacy (DP for short) is a widely-used technology to defend against the above attacks with a rigorous privacy guarantee~\cite{auditingDPSGD}, which is different from the protections provided by current heuristic methods~\cite{memguard, li2021membership}. Meanwhile, a recent user study~\cite{xiong2020towards} shows that users prefer to share their data when a rigorous privacy guarantee is provided. Hence, DP could be a more significant defense than existing heuristic defenses.  
However, the DP noise would lead to significant accuracy loss of trained models, especially for deep neural network models~\cite{tramer2021differentially, deepdp}. In addition, even though there have been many studies on the efficiency optimization of MPL~\cite{mohassel2018aby3,cryptGPU}, there is still a huge efficiency gap between plaintext training and secure training. Therefore, it is highly challenging to balance the 3-way trade-off between privacy, efficiency, and accuracy in the secure training of MPL. Concretely, (1) directly combining the plaintext data from all parties to efficiently train an accurate model would bring severe privacy concerns; (2) training an accurate model with existing MPL frameworks can partially protect privacy, but it requires a large amount of time to complete the training process; (3) while integrating DP into MPL frameworks without further optimizations can resolve the privacy concerns, the accuracy loss led by DP would cancel out the accuracy gain of combining multiple parties' data.

In this paper, we are motivated to resolve the above issue by proposing \wqruan{a solution, referred to as PEA (Private, Efficient, Accurate), which consists of a secure differentially private stochastic gradient descent (DPSGD for short) protocol and two optimization methods.} First, by designing a secure inverse of square root protocol, we enforce the DPSGD algorithm~\cite{deepdp,song2013stochastic, bassily2014private} in secret sharing-based MPL frameworks to enhance their privacy protection.
Second, to reduce the accuracy loss led by DP and improve the efficiency of MPL, we propose two optimization methods: (1) extracting the features of input data with data-independent feature extractors; (2) initializing the global model by aggregating local models trained on local data from parties. 
With the first optimization method, parties only need to securely train a differentially private simple model on extracted features, which can achieve similar, even higher accuracy than differentially private complex models, such as Resnet with tens of layers~\cite{7780459},  trained on raw data. In addition, the second optimization method initializes a relatively accurate global model for the secure training process, thus significantly speeding up the convergence of the model training process.          

We summarize our main contributions as follows:
\begin{itemize}
    \item We propose the secure DPSGD protocol to enforce the DPSGD algorithm in secret sharing-based MPL frameworks, thus defending against the membership inference attacks on machine learning models (Section~\ref{secure DPSGD}).
    
    \item  We propose two optimization methods to reduce the accuracy loss brought by DP and improve the efficiency of MPL frameworks: (1) the data-independent feature extraction method (Section~\ref{subsec:separate});  (2) the local data-based global model initialization method (Section~\ref{subsec:initialization}). 
    

    \item We implement \wqruan{PEA} in two open-source MPL frameworks, namely  \texttt{TF-Encrypted}~\cite{TF-Encrypted} and \texttt{Queqiao}\footnote{\url{https://github.com/FudanMPL/SecMML}}. We refer to the differentially private versions of \texttt{TF-Encrypted} and \texttt{Queqiao} as \tfencrypted and \queqiao respectively. \wqruan{The experimental results on three widely-used datasets show that our proposed feature extraction method and model initialization method can significantly improve the efficiency of MPL frameworks.} In particular, under the LAN setting, when we set $\epsilon$ as 2, \tfencrypted and \queqiao can train a differentially private classification model for CIFAR-10 with an accuracy of 88\% within 7 and 55 minutes, respectively. While \texttt{CryptGPU}~\cite{cryptGPU}, one state-of-the-art MPL framework, 
   requires more than 16 hours (about 137$\times$ and 17$\times$ of ours respectively) to train a non-private deep neural network model for CIFAR-10 with the same accuracy (Section~\ref{experiment}). \wqruan{Therefore, with our proposed PEA, \tfencrypted and \queqiao can balance privacy, efficiency, and accuracy in the secure model training of MPL.} 
\end{itemize}

\vspace{-1mm}
\section{Overview}~\label{sec:overview}
\begin{figure}[ht]
    \centering
    \includegraphics[scale=0.38]{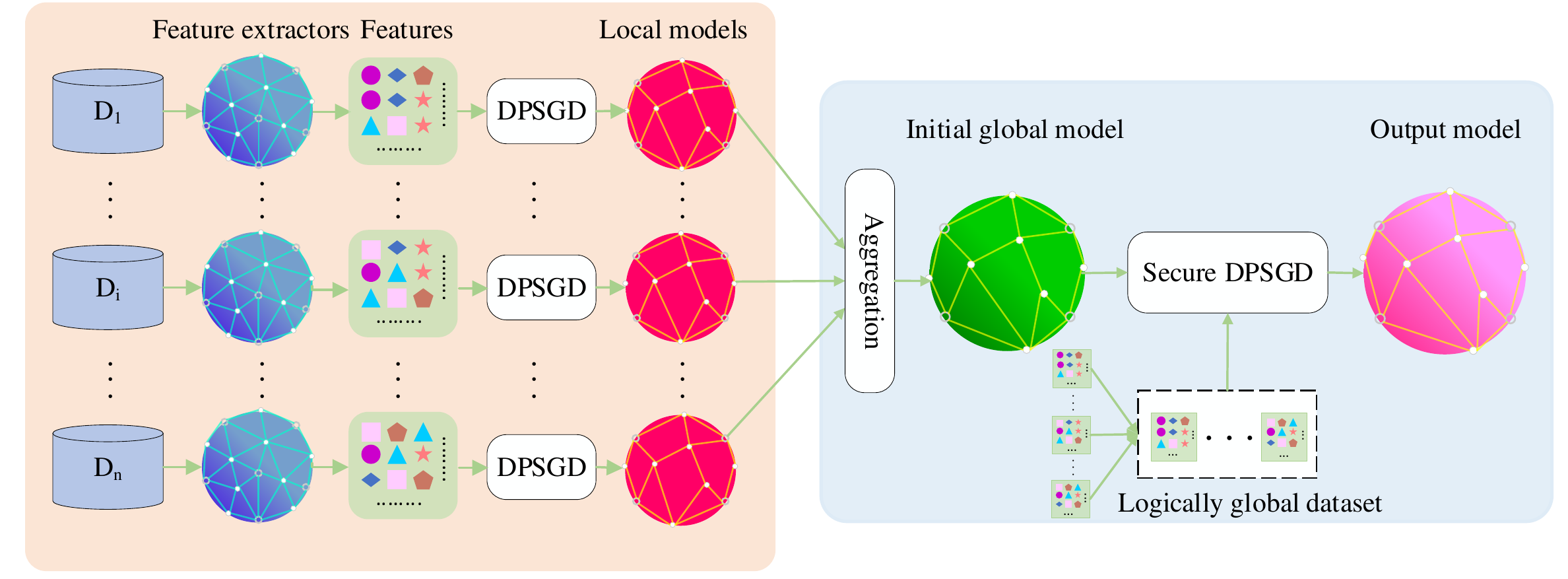}
    \caption{An overview of \wqruan{PEA}. The computations in the orange (left) frame are performed locally on plaintext data, and computations in the blue (right) frame are performed globally on secret shared data. During the global computation, all local datasets are merged as the logically global dataset through using SMPC protocols.}
    \label{fig:overview}
\end{figure}

As is shown in Figure~\ref{fig:overview}, we present an overview of \wqruan{PEA}, which works in the training process with the support of SMPC protocols. In the following parts, we introduce the motivation and design of our proposed protocol and optimization methods.

\vspace{-1mm}
\subsection{Secure DPSGD Protocol}
In the multi-party setting, local differential privacy~\cite{foundation_dp}, which has no dependence on a trusted server, is a straightforward method to protect privacy. However, the accuracy loss caused by local differential privacy is one or more orders of magnitude larger than that caused by central DP~\cite{kasiviswanathan2011can}, which requires a trusted server to handle all data.
Benefit from the security guarantee provided by SMPC protocols, even though in the multi-party setting, MPL can simulate a trusted server that trains the model on all data from distributed parties. 

According to the above analysis, we leverage the DPSGD algorithm~\cite{deepdp,song2013stochastic,bassily2014private} in the multi-party setting to protect models trained by MPL frameworks. DPSGD adds noise sampled from Gaussian distribution to gradient vectors with bounded $L_2$ norms to protect training data. We thus design a secret sharing-based inverse of square root protocol to clip $L_2$ norms of gradient vectors and utilize the distributed noise generation protocol proposed by Dwork et al.~\cite{10.1007/11761679_29} to generate random Gaussian noise securely.

\vspace{-2mm}
\subsection{Efficiency and Accuracy Optimization Methods}
Existing MPL frameworks~\cite{DBLP:conf/sp/MohasselZ17,mohassel2018aby3,wagh2020falcon,cryptGPU} try to securely train end-to-end deep neural network models (i.e. training neural network models that combine feature extraction and classification). However, combining DP with deep neural network models could result in significant accuracy loss~\cite{deepdp, tramer2021differentially}. Furthermore, under the computation and communication complexity of current SMPC protocols, it is difficult, if not impractical, to securely train an end-to-end deep neural network model. We elaborate on these two arguments in Section~\ref{subsec:separate}. 

Instead of training an end-to-end neural network model, we let each party first apply data-independent feature extractors, including foundation models~\cite{foundation_model} (e.g. BERT~\cite{BERT}) and feature extraction rules, to extract features of input datasets locally. Consequently, all parties collaboratively train shallow models (e.g. linear models or neural network models with a few layers) on extracted features. In this way, parties only need to perform secure computation and add noise in the shallow model training process. As a result, the accuracy loss and communication overhead can be significantly reduced.

Furthermore, when we train a machine learning model, starting from an accurate initial model would significantly improve the convergence speed. Our main idea is to let each party train a local model on his or her local data with DPSGD. Subsequently, parties securely aggregate their local models with SMPC protocols to initialize a global model. By performing parts of the computation of the training process locally, parties can therefore significantly improve the efficiency of MPL.

\vspace{-2mm}
\section{Preliminaries}~\label{preliminaries}
We first describe the security model used in this paper. After that, we introduce background knowledge of SMPC and differentially private machine learning. For clarity purposes, we list notations used in the paper in Table~\ref{tab:notations}.

\begin{table}[ht]
\centering
\caption{Notations used in this paper.}
\scalebox{0.85}{
\begin{tabular}{|c|l|}
\hline
$P_i$         & Party $i$ participating in MPL                                                      \\ \hline
$m$           & The number of parties involved in MPL                                               \\ \hline
$\share{x}_i$ & One share of a secret $x$ held by $P_i$                                             \\ \hline
$k$           & The number of bits used to represent a fixed-point number \\ \hline
$f$           & The bit number of fixed-point number's fraction part                                \\ \hline
$p$           & The dimension of gradient vectors                                                   \\ \hline
$n$           & The number of data samples                                                          \\ \hline
$\theta$      & Trained model                                                                       \\ \hline
$C$           & The gradient clipping bound                                                         \\ \hline
$\eta$        & Learning rate                                                                       \\ \hline
$B$           & Batch size                                                                          \\ \hline
\end{tabular}
}
\label{tab:notations}
\end{table}

\vspace{-1mm}
\subsection{Security Model}
In this paper, we consider the semi-honest security model, which is widely used in MPL frameworks~\cite{mohassel2018aby3,DBLP:conf/sp/MohasselZ17}. In the model, each party tries to infer private information from all received messages but will follow the steps of protocols.  The semi-honest model should be a reasonable setting because the primary goal of MPL is to bridge the data sources separated by strict privacy protection regulations, and parties are willing to share their data. \wqruan{Furthermore, following previous studies~\cite{mohassel2018aby3,cryptGPU},  we assume an honest majority, i.e. less than half of parties would collude. In summary, PEA guarantees the security of data during the model training process in the semi-honest setting with an honest majority, and provides DP  guarantees to defend against membership inference attacks on output models.}

\vspace{-1mm}
\subsection{Secure Multi-party Computation}\label{pre:smpc}
SMPC enables multiple parties to collaboratively execute a function without leaking any information of each party's input except the computation results. There are several technical routes~\cite{ref_yao,ref_damgard,yi2014homomorphic} to implement SMPC, including secret sharing~\cite{ref_damgard}. The main idea of secret sharing is to divide a private value $x$ into $m$ shares, where $m$ is the number of parties. Subsequently, parties perform computations of a target function on the shares and finally reconstruct the computation results as the output.  The major MPL frameworks are based on linear secret sharing protocols (i.e. additive secret sharing~\cite{ref_gmw} and Shamir's secret sharing~\cite{ref_bgw}) for their high efficiency of arithmetic computations~\cite{mohassel2018aby3, DBLP:conf/sp/MohasselZ17,wagh2020falcon, cryptGPU}. 

We construct the secure DPSGD protocol based on linear secret sharing protocols.  Because different secret sharing schemes have different sharing semantics, we use the primitives that are supported by all linear secret sharing protocols to construct our proposed secure DPSGD protocol. Thus our proposed protocol can be applied to linear secret sharing-based MPL frameworks, including \texttt{ABY3}~\cite{mohassel2018aby3} and \texttt{Queqiao}. 

We hereby introduce the fixed-point number representation method and cryptographic primitives used in this paper. As these primitives are used as black-boxes, we only introduce details of secret sharing and reconstruction. For other primitives, we present the functionality of these primitives and refer details to the related literature~\cite{div2mp,10.1007/978-3-642-15317-4_13,ref_bgw,ref_gmw}.  Note that in our computation process,  all parties are usually equal to each other. This feature is different from federated learning~\cite{yang2019federated}, where there is usually a central server among parties. 

\noindent\textbf{Fixed-point Number Representation.} Original linear secret sharing protocols~\cite{ref_bgw,ref_gmw} only support secure computations on integers. However, the model training process often involves floating-point numbers. Therefore, we encode a fixed-point number with $f$-bit precision by mapping a fixed-point number $\widetilde{x}$ to an integer $\overline{x}$ such that $\overline{x} = \widetilde{x} * 2^f$. In order to prevent the precision explosion brought by fixed-point number multiplication (i.e. $\overline{z} = \overline{x} * \overline{y} = \widetilde{x} * \widetilde{y} * 2^{2f}$), we apply $\mathsf{Trunc}$ protocol~\cite{div2mp} to truncate $\overline{z}$ as $\overline{z}'$ such that $\overline{z}' =\widetilde{x} * \widetilde{y} * 2^{f} $. $\mathsf{Trunc}$ receives three parameters: $x, k, q$, where $x$ is the data to be truncated,  $k, q$ indicate the number of bits to represent the input data and the number of bits to be truncated respectively.

\noindent\textbf{Secret Sharing and Reconstruction.}  For additive secret sharing~\cite{ref_gmw}, given a secret $x$, we first generate $m-1$ random number $\share{x}_1, \share{x}_2, \cdots, \share{x}_{m-1}$, and let $\share{x}_m = x  - \sum_{i=1}^{m-1}\share{x}_i$. $P_i$ holds $\share{x}_i$. To reconstruct $x$, all parties exchange $\share{x}_i$ and add them together to obtain $x$. For Shamir's secret sharing~\cite{ref_bgw}, given a secret $x$, we first generate a polynomial $f(y) = a_0 + a_1y+a_2y^2+...+a_ty^t$, where $a_0=x, a_1,\cdots,a_t $ are random coefficients, and $t+1$ is the least number of parties to reconstruct the secret; then we construct $m$ points out of it with indexes $1\leq i \leq m$. These $f(i)$ (i.e. $\share{x}_i$) are exactly the shares of $x$. To reconstruct $x$, all or some of all parties exchange $\share{x}_i$ and reconstruct the polynomial $f(y)$ with the Lagrange interpolation to obtain $x$.

\noindent\textbf{Unary Operations.} Given one share $\share{x}_i$ held by $P_i$:
\begin{itemize}
    \item $\mathsf{Bit\_Dec}$ outputs $\share{x_0}_i, \share{x_1}_i, \cdots, \share{x_{k-1}}_i$ such that $x = \sum_{t = 0}^{k-1}x_t*2^t$.
    
    \item $\mathsf{Mod2}$ outputs $\share{t}_i$ for $P_i$ such that $ t= x\mod2$. 
\end{itemize}

\noindent \textbf{Binary Operations.} Given two shares $\share{x}_i$, $\share{y}_i$ held by $P_i$:

\begin{itemize}
    \item $\mathsf{Addition}$ outputs $\share{z}_i$ for $P_i$ such that $z = x+y$.
    \item $\mathsf{Multiplication}$ outputs $\share{z}_i$ for $P_i$ such that $z = x*y$. 
    \item $\mathsf{Comparison}$ outputs $\share{z}_i$ for  $P_i$ such that $z = x > y  ?  0 : 1$.
\end{itemize}

\noindent\textbf{Multiple-Input Operations.} Given a sequence of shares $\share{x_0}_i, \share{x_1}_i, \cdots, \share{x_{k-1}}_i$ held by $P_i$:
\begin{itemize}
    \item $\mathsf{SufOr}$ outputs $\share{c_0}_i, \share{c_1}_i, \cdots, \share{c_{k-1}}_i$ such that $c_0 = x_0 \vee x_1 \vee \cdots \vee x_{k-1}, \cdots, c_{k-1} = x_{k-1} $.
    
    \item $\mathsf{PreMulC}$ output $\share{c_0}_i, \share{c_1}_i, \cdots, \share{c_{k-1}}_i$ such that $c_0 = x_0, c_1 = x_0*x_1, \cdots, c_{k-1} = x_0*x_1*\cdots*x_{k-1}$.
\end{itemize}

\vspace{-1mm}
\subsection{Differentially Private Machine Learning}\label{sec:differential_privacy}
\begin{myDef}
\textbf{\begin{math}(\epsilon,\delta)\end{math}-Differential Privacy}~\cite{foundation_dp}. For a random mechanism M whose input is \begin{math}D \end{math} and outputs \begin{math}r \in R^p\end{math}, we say M is  \begin{math}(\epsilon,\delta)\end{math}-differential \ private if for any subset \begin{math}S \subseteq  R^p\end{math},
\begin{center}
\scalebox{0.9}{
    \begin{math}
    Pr(M(D) \in S) \leq e^{\epsilon} \cdot Pr (M(D') \in S) + \delta
    \end{math}
}
\end{center}
\end{myDef}
\noindent where $D$ and $D'$ \begin{math}\end{math} are neighboring datasets, i.e. we can obtain $D'$ by deleting or adding one data sample of $D$. If $\delta$ is set as 0, we say $M$ is \begin{math}\epsilon\end{math}-differentially private.  $\epsilon$ is also called privacy budget that represents the privacy loss of a DP mechanism. When multiple DP mechanisms are performed on the same dataset simultaneously, the total privacy budget is estimated by composition theorems, e.g. sequential composition theorem~\cite{foundation_dp} and parallel composition theorem~\cite{parallel_composition}.
\begin{theorem}~\label{sequential_composition}
\textbf{Sequential Composition Theorem~\cite{foundation_dp}.}
For n mechanisms $M_1, M_2, \cdots$ $, M_n$ that access the same dataset D, if each $M_i$ satisfies ($\epsilon_i$, $\delta_i$)-DP, the combination of their outputs satisfies ($\epsilon$, $\delta$)-DP with $\epsilon =\epsilon_1 + \epsilon_2 + \cdots + \epsilon_n$, $\delta = \delta_1 + \delta_2 + \cdots + \delta_n$.
\end{theorem}
\begin{theorem}~\label{original_parallel_composition}
\textbf{Parallel Composition Theorem~\cite{parallel_composition}.} For n mechanisms $M_1, M_2, \cdots$ $, M_n$  whose inputs are disjoint datasets $D_1, D_2, \cdots, D_n$, if $M_i$ satisfies $\epsilon_i$-DP, the combination of them satisfies $\epsilon$-DP for the dataset  $D = D_1 \cup D_2 \cup \cdots \cup D_n$ with $\epsilon = \mathit{Max}(\epsilon_1, \epsilon_2, \cdots, \epsilon_n)$.
\end{theorem}
Based on Theorem~\ref{original_parallel_composition}, we elaborate the parallel composition theorem for ($\epsilon, \delta$)-DP in the Lemma~\ref{parallel_composition}, whose proof is shown in Appendix~\ref{Missing Proof}.

\begin{lemma}~\label{parallel_composition}
\textbf{Parallel Composition Theorem for ($\epsilon, \delta$)-Differential Privacy.} For n mechanisms $M_1, M_2, \cdots$ $, M_n$  whose inputs are disjoint datasets $D_1, D_2, \cdots, D_n$, if $M_i$ satisfies ($\epsilon_i, \delta_i$)-DP, the combination of them satisfies ($\epsilon$, $\delta$)-DP  for the dataset  $D = D_1 \cup D_2 \cup \cdots \cup D_n$ with $\epsilon = \mathit{Max}(\epsilon_1, \epsilon_2, \cdots, \epsilon_n)$, $\delta = \mathit{Max}(\delta_1, \delta_2, \cdots, \delta_n)$.
\end{lemma}
\noindent In addition to composition theorems, another important property of DP is post-processing immunity~\cite{foundation_dp}. That is, performing data-independent computations over the differentially private output will not influence its privacy guarantee. 

\floatname{algorithm}{Algorithm}
\begin{algorithm}
\small
\caption{DPSGD~\cite{deepdp,song2013stochastic,bassily2014private}}
\label{DPSGD}
\begin{algorithmic}[1]
\REQUIRE  Data set $D = \{d_1,\cdots, d_n\}$, loss function: $\ell(\theta; d_i)$, privacy parameters: ($\epsilon, \delta$), number of iterations: $T$, batch size: $B$, gradient clipping bound $C$, learning rate function: $\eta$;
\ENSURE Trained model $\theta_T$;
\STATE Randomly initialize model $\theta_1$;
\FOR{$t=1 \ to \ T $}
\STATE $s_1,\cdots,s_B \leftarrow$ Sample $B$ samples uniformly from $D$;
\STATE For each $s_i$, compute $\nabla\ell(\theta;s_i)$;
\STATE \textbf{Clip gradient}
\STATE$\nabla\ell(\theta;s_i) \leftarrow \nabla\ell(\theta;s_i) * min(1, \frac{C}{\left\| \nabla\ell(\theta;s_i) \right\|_2})$  ;
\STATE \textbf{Add noise and Descent} 
\STATE $b_t \sim \mathcal{N}(0, \sigma^2I_{p\times p})$, where $\sigma \geq C\frac{2\sqrt{T\log\frac{1}{\delta}}}{\epsilon}$;
\STATE $\theta_{t+1} = \theta_t - \frac{\eta(t)}{B}[\sum_{i=1}^{B}\nabla\ell(\theta;s_i) + b_t]$;
\ENDFOR
\STATE Output $\theta_{T+1}$;
\end{algorithmic}
\end{algorithm}
\floatname{algorithm}{Protocol}
 
\setcounter{algorithm}{0}

We then introduce the DPSGD algorithm, whose details are shown in Algorithm~\ref{DPSGD}. In order to securely perform DPSGD, besides gradient computation and descent that have been implemented in existing MPL frameworks, there are two key steps to be completed: (1) securely clipping the gradient vectors such that their $L_2$ norms are all lower than or equal to a given constant $C$; (2) securely generating the random noise vectors sampled from Gaussian distribution. 



\section{Secure DPSGD}~\label{secure DPSGD}
In this section, we first introduce the secure inverse of square root and secure Gaussian noise generation protocols. Then we combine these two protocols to design the secure DPSGD protocol. 

\subsection{Inverse of Square Root} \label{Sec:Inverse}
In order to clip the $L_2$ norm of one gradient vector $\textbf{g}$, parties need to compute the inverse of $\left \| \textbf{g} \right\|_2$, i.e. $\frac{1}{\sqrt{g_1^{2} + g_1^{2} + \cdots + g_p^{2} }}$. As this computation involves two functions of square root and division, which are both non-linear, directly implementing them with secret sharing protocols implies huge communication overhead. To reduce the overhead, a feasible way is to use polynomial to approximate the non-linear functions. Lu et al.~\cite{inverse_root_ccs} proposed to transform the original secret value $\share{x}_i$ to a form $\share{x'*2^{exp}}_i$, where $ 0.25 \leq x' \leq 0.5$. Afterwards, they separately compute the $\share{\frac{1}{\sqrt{x'}}}_i$ and  $\share{2^{-\frac{exp}{2}}}_i$. Finally, they obtain $\share{\frac{1}{\sqrt{x}}}_i$ by multiplying $\share{\frac{1}{\sqrt{x'}}}_i$ and $\share{2^{-\frac{exp}{2}}}_i$.  \wqruan{However, their method has an issue, i.e. the error of their polynomial approximation might destroy the privacy guarantee of DP.} That is, if true $\frac{C}{\left\| \nabla\ell(\theta;s_i) \right\|_2}$ is slightly smaller than 1 (i.e. $\left\| \nabla\ell(\theta;s_i) \right\|_2 > C$) and approximated $\frac{C}{\left\| \nabla\ell(\theta;s_i) \right\|_2}$ is slightly higher than 1 (i.e. $\left\| \nabla\ell(\theta;s_i) \right\|_2^{\mathit{approx}} < C$), the gradient vector will not be clipped according to Line 6 of Algorithm~\ref{DPSGD}, and its $L_2$ norm will exceed $C$. Consequently, the privacy guarantee of DP is destroyed.
\begin{algorithm}
\small
\caption{ Inverse of Square Root }
\label{alg:secure_inversesqrt}
\begin{algorithmic}[1]
\REQUIRE  $P_i$ holds the shares of the input $\share{x}_i$;
\ENSURE $P_i$ obtains the shares of the output $\share{y}_i$, such that $y = \frac{1}{\sqrt{x}}$;
\STATE \quad \textbf{Data preparation and transformation}
\STATE $\{\share{x_t}_i\}_{t=0}^{k-1} \leftarrow \mathsf{Bit\_Dec}(\share{x}_i)$;
\STATE $\{\share{c_t}_i\}_{t=0}^{k-2} \leftarrow \mathsf{SufOr}(\{\share{x_t}_i\}_{t=0}^{k-2})$ ;
\STATE $\share{b}_i = 1 + \sum_{t=0}^{k-2}2^{k-2-t}(1-\share{c_t}_i)$;
\STATE $\share{x'}_{i} = \mathsf{Trunc}(\share{b}_i * \share{x}_i, k, k-f-1)$;
\STATE $\share{\mathit{exp}}_i \leftarrow \sum_{t=0}^{k-2}\share{c_t}_i - f$;
\STATE \quad \textbf{Compute the square root of $2^{-\mathit{exp}}$};
\STATE $\share{lsb}_i = \mathsf{Mod2}(\share{\mathit{exp}}_i)$;
\STATE $\share{\frac{\mathit{exp}}{2}}_i = \mathsf{Trunc}(\share{\mathit{exp}}_i, k, 1)$;
\STATE $\{\share{e_t}_i\}_{t=0}^{k-1} =  \mathsf{Bit\_Dec}(\share{f- \frac{\mathit{exp}}{2}}_i)$;
\FOR{$t=0 \ to \ k-1 $}
\STATE $\share{e_t}_i = (1 + \share{e_t}_i * (2^{2^{t}} - 1))$;
\ENDFOR
\STATE $
\{\share{c_t}_i\}_{t=0}^{k-1}= \mathsf{PreMulC}(\{\share{e_t}_i\}_{t=0}^{k-1})$;
\STATE $\share{2^{f- \frac{\mathit{exp}}{2}}}_i =  \share{c_{k-1}}_i$;
\STATE $\share{2^{f- \frac{\mathit{exp}}{2}-\frac{1}{2}}}_i = \mathsf{Trunc}(\share{2^{f- \frac{\mathit{exp}}{2}}}_i*2^{f-\frac{1}{2}}, k, f)$;
\STATE $\share{2^{f- \frac{\mathit{exp}}{2}}}_i = \share{2^{f- \frac{\mathit{exp}}{2}-\frac{1}{2}}}_i*\share{lsb}_i + \share{2^{f- \frac{\mathit{exp}}{2}}}_i * (1 - \share{lsb}_i)$;
\STATE \quad \textbf{Compute approximated polynomial}
\STATE $\share{x'_{\mathit{approx}}}_i = \mathsf{Trunc}(0.8277*2^{f}*\share{x'}_i, k, f) $;
\STATE $\share{x'_{\mathit{approx}}}_i = \mathsf{Trunc}((\share{x'_{\mathit{approx}}}_i - 2.046*2^{f})*\share{x'}_i, k, f) $;
\STATE $\share{x'_{\mathit{approx}}}_i=\share{x'_{\mathit{approx}}}_i + 2.223*2^{f} - 0.0048*2^{f} $;
\STATE \quad \textbf{Compute and output final result}
\STATE $\share{y}_i= \mathsf{Trunc}(\share{x'_{\mathit{approx}}}_i *\share{2^{f- \frac{\mathit{exp}}{2}}}_i, k, f) $;
\STATE $P_i$ obtains $\share{y}_i$;
\end{algorithmic}
\end{algorithm}

In order to resolve the above issue, we propose the inverse of square root protocol based on linear secret sharing protocols and prove an error bound of the approximated polynomial to keep the rigorous privacy guarantee of DP. As is shown in Protocol~\ref{alg:secure_inversesqrt}, we describe each step of the inverse of square root protocol as follows:

\noindent$\bullet$ \textbf{Step 1.} (Line 2-6)   $P_i$ first decomposes $\share{x}_i$ as binary encoding $\{\share{x_t}_i\}_{t=0}^{k-1}$ and performs $\mathsf{SufOr}$ on $\{\share{x_t}_i\}_{t=0}^{k-2}$ so that the bits after the first 1 in the binary encoding are all set as 1. Finally, $P_i$ transforms $\share{x}_i$ to $\share{x'}_i $ and obtains the $\share{\mathit{exp}}_i$ such that $x' \in [0.5, 1)$ and $x = x' * 2^{\mathit{exp}}$.

\noindent$\bullet$ \textbf{Step 2.} (Line 8-17) $P_i$ computes the fixed-point representation of  $\share{2^{-\frac{\mathit{exp}}{2}}}_i$, i.e. $\share{2^{f-\frac{\mathit{exp}}{2}}}_i$. First, $P_i$ obtains the least significant bit (i.e. $\share{lsb}_i$) of $\mathit{exp}$, which indicates the parity of $\mathit{exp}$ through applying $\mathsf{Mod2}$.  Subsequently, $P_i$ computes $\share{2^{f-\frac{\mathit{exp}}{2}}}_i$ and $\share{2^{f - \frac{\mathit{exp}}{2} -\frac{1}{2}}}_i$ respectively, where $\frac{\mathit{exp}}{2} = \frac{\mathit{exp}-1}{2}$ if $\mathit{exp}$ is odd. Finally, $P_i$ obtain $\share{2^{f-\frac{\mathit{exp}}{2}}}_i$ by choosing the correct value based on the value of $\share{lsb}_i$.

\noindent$\bullet$ \textbf{Step 3.} (Line 19-21) $P_i$ approximates $\frac{1}{\sqrt{x'}}$ for $x' \in [0.5, 1)$ with polynomial $0.8277x'^2 - 2.046x' + 2.223$, which is obtained through using the $\mathsf{polyfit}$ function of NumPy~\footnote{\url{https://numpy.org/doc/stable/reference/generated/numpy.polyfit.html}}. Next, $P_i$ subtracts the approximated value with error bound 0.0048 to ensure that the approximated value is strictly smaller than the true value, thus keeping the rigorous privacy guarantee. 

\noindent$\bullet$ \textbf{Step 4.} (Line 23-24) $P_i$ multiplies the approximated value $\share{x'_{\mathit{approx}}}_i$ with $\share{2^{f - \frac{\mathit{exp}}{2}}}_i$ to obtain the output $\share{y}_i$.

We bound the difference between our approximated polynomial and true output as follows. 
\begin{lemma} \label{lemma:error_bound}
For any $x'$ $ \in [0.5, 1),  (0.8277x'^2-2.046x'+2.223 - 0.0048) -\frac{1}{\sqrt{x'}} < 0$.
\end{lemma}
We show the proof of Lemma~\ref{lemma:error_bound} in Appendix~\ref{Missing Proof}. When the approximated $\frac{1}{\sqrt{x'}}$ is smaller than the true $\frac{1}{\sqrt{x'}}$, the approximated $\frac{1}{\sqrt{x}}$ is smaller than the true $\frac{1}{\sqrt{x}}$. Therefore, the approximated $\frac{C}{\left\| \nabla\ell(\theta;s_i) \right\|_2^{\mathit{approx}}}$ (i.e. $\frac{1}{\sqrt{\sum_{j=0}^{p-1}\nabla\ell(\theta;s_i)_j^2}} *C$ ) is smaller than the true value. Consequently, according to Algorithm~\ref{DPSGD}, we can guarantee that gradient vectors whose $L_2$ norms are larger than $C$ (i.e. $\frac{C}{\left\| \nabla\ell(\theta;s_i) \right\|_2} < 1$ ) are all clipped. Finally, we can obtain the DP guarantee of true gradient vectors. We further analyze the security and communication complexity of Protocol~\ref{alg:secure_inversesqrt} in Appendix~\ref{sec:analysis_inversesquareroot}.



\vspace{-1mm}
\subsection{Noise Generation}~\label{Sec:Noise_Gen}
We refer to the noise generation protocol proposed by Dwork et al.~\cite{10.1007/11761679_29} to securely generate the Gaussian noise with mean 0 and standard deviation $\sigma$. The details are shown in Protocol~\ref{alg:secure_noise_generation}. Benefit from the additivity of Gaussian distribution, the sum of independent Gaussian noises from parties is the target Gaussian noise. According to \cite{10.1007/11761679_29}, each party generates a Gaussian noise with variance $\frac{3}{2m}\sigma^2I_{p\times p}$ to preserve Byzantine robustness of Protocol~\ref{alg:secure_noise_generation}. \wqruan{Byzantine robustness means that if not more than 1/3 of parties fail, parties can generate the noise that is random enough. Without the Byzantine robustness here, when part of parties fails during the noise generating process, the generated noise may not be random enough to provide privacy protection, which is the main goal of our proposed PEA.} Note that this is not the property of the end-to-end secure DPSGD protocol. Meanwhile, as the noise generation process only requires secure addition, its security is trivially guaranteed by the security of secure addition. As to the communication complexity, Protocol~\ref{alg:secure_noise_generation} can be completed in one communication round with $O(mpk)$ bits message transfer. Besides, as the noise generation process is data-independent, it can be generated offline and has no impact on the efficiency of the online training process.

\begin{algorithm}[ht]
\small
\caption{ Secure Gaussian Noise Generation~\cite{10.1007/11761679_29}}
\label{alg:secure_noise_generation}
\begin{algorithmic}[1]
\REQUIRE  The standard deviation of the target Gaussian distribution $\sigma$, the dimension of noise $p$;
\ENSURE The shares of a Gaussian noise $\share{b}_i$ such that $b \sim \mathcal{N}(0, \sigma^2I_{p\times p})$;
\STATE $P_i$ samples a Gaussian noise $b_i  \sim \mathcal{N}(0, \frac{3}{2m}\sigma^2I_{p\times p})$;
\STATE $P_i$ sends the shares of $\share{b_i}_{j \neq i}$ to other parties  $P_{j \neq i}$;
\STATE $P_i$ summarizes the noise shares from other parties to obtain the result $\share{b}_i = \sum_{j=1}^{m}\share{b_j}_i $;
\end{algorithmic}
\end{algorithm}

\vspace{-2mm}
\subsection{Secure DPSGD Protocol}~\label{subsec:secure_DPSGD}
With Protocol~\ref{alg:secure_inversesqrt} and Protocol~\ref{alg:secure_noise_generation}, we then introduce the secure DPSGD protocol, as is shown in Protocol~\ref{alg:secure_dp_sgd}. First, parties securely compute gradient vectors. Second, parties clip the $L_2$ norms of the gradient vectors through calculating the inverse of $L_2$ norms of the gradient vectors with Protocol~\ref{alg:secure_inversesqrt} and storing the comparison results between them with 1 in $\mathit{is}\_\mathit{clip}$. After that, parties update the gradient vectors according to the values of $\mathit{is}\_\mathit{clip}$.  Subsequently, all parties collaboratively generate Gaussian noises with a pre-defined standard deviation $\sigma$ through calling Protocol~\ref{alg:secure_noise_generation} and apply them to perturb the gradient vectors. Finally, parties update model parameters $\theta$ by following the regular gradient descent process. Note that the model parameters are kept in the secret shared form through the whole secure DPSGD process.
We analyze the security and communication complexity of Protocol~\ref{alg:secure_dp_sgd} in Appendix~\ref{sec:analysis_secure_dpsgd}.

\noindent\textbf{Privacy Guarantee.} We show the privacy guarantee of Protocol~\ref{alg:secure_dp_sgd} in Theorem~\ref{theo:secure_dp_sgd} (its proof is shown in Appendix~\ref{Missing Proof}). 
\begin{theorem} \label{theo:secure_dp_sgd}
  For any $\epsilon \leq 2\log(1/\delta)$ and $\delta \in (0, 1)$,  Protocol~\ref{alg:secure_dp_sgd} satisfies ($\epsilon, \delta$)-DP.
\end{theorem}

\noindent\textbf{Random Sampling of Minibatches. } Protocol~\ref{alg:secure_dp_sgd} amplifies the privacy guarantee of the training process by randomly sampling minibatches from the logically global dataset. \wqruan{We apply the resharing-based oblivious shuffling protocol~\cite{laur2011round}, which can securely shuffle the logically global dataset while keeping the new permutation of data samples invisible to parties, to implement the random sample operation. With 
the shuffling protocols, we can securely sample minibatches to meet the requirement of DPSGD. Note that the oblivious shuffling protocol assumes an honest majority. } 





\begin{algorithm}
\caption{Secure DPSGD}
\small
\label{alg:secure_dp_sgd}
\begin{algorithmic}[1]
\REQUIRE  $P_i$ holds a local dataset $D_i$, loss function: $\ell(\theta; d_i)$, privacy parameters: ($\epsilon, \delta$), number of iterations: $T$, batch size: $B$, gradient clipping bound $C$, learning rate function: $\eta$;
\ENSURE $P_i$ obtains the shares of the trained model $\share{\theta_T}_i$;
\STATE $P_i$ generates and sends the shares of its own dataset $\share{D_i}_{j \neq i}$ to $P_{j \neq i}$. Thus, $P_i$ holds the share of the global dataset $\share{D}_i = \{\share{D_1}_i, \share{D_2}_i, \cdots, \share{D_m}_i\}$;
\STATE Initializing $\share{\theta_1}_i$ with Protocol~\ref{alg:global_model_initialization};
\FOR{$t=1 \ to \ T $}
\STATE $\share{s_1}_i,\cdots,\share{s_B}_i \leftarrow$ parties collaboratively sample $B$ samples from $D$;
\STATE For each $\share{s_j}_i$, compute $\share{\nabla\ell(\theta_t;s_j)}_i$;
\STATE \textbf{Clip gradient}
\FOR{each $\share{\nabla\ell(\theta_t;s_j)}_i$}
\STATE $\share{x_j}_i = \mathsf{Trunc}(\share{\nabla\ell(\theta_t;s_j)}_i * \share{\nabla\ell(\theta_t;s_j)^T}_i, k, f )$;
\STATE Compute the inverse square root of $\share{x_j}_i$ (i.e. $\share{\frac{1}{\left\|\nabla\ell(\theta_t;s_j)\right\|_2^{\mathit{approx}}}}_i$) with Protocol~\ref{alg:secure_inversesqrt};
\STATE $\share{is\_clip}_i = \mathsf{Comparison}( \share{\frac{C}{\left\|\nabla\ell(\theta_t;s_j)\right\|_2^{\mathit{approx}}}}_i, 1)$;
\STATE $\share{\nabla\ell(\theta_t;s_j)_{clipped}}_i =  \mathsf{Trunc}(\share{\nabla\ell(\theta_t;s_j)}_i*\share{\frac{C}{\left\|\nabla\ell(\theta_t;s_j)\right\|_2^{\mathit{approx}}}}_i,k,f)$;
\STATE$\share{\nabla\ell(\theta_t;s_j)}_i \leftarrow \share{\nabla\ell(\theta_t;s_j)}_i * (1- \share{is\_clip}_i) +  \share{\nabla\ell(\theta_t;s_j)_{clipped}}_i*\share{is\_clip}_i$;
\ENDFOR
\STATE \textbf{Add noise and Descent} 
\STATE Generate noise $\share{b_{t}}_i$ such that $b_{t} \sim \mathcal{N}(0, \sigma^2I_{p\times p})$ with Protocol~\ref{alg:secure_noise_generation}, where $\sigma \geq C\frac{2\sqrt{T\log\frac{1}{\delta}}}{\epsilon}$;
\STATE $\share{update}_i = \mathsf{Trunc}(\frac{\eta}{B}[\sum_{j=1}^{B}\share{\nabla\ell(\theta_t;s_j)}_i + \share{b_t}_i],k,f)$;
\STATE $\share{\theta_{t+1}}_i = \share{\theta_t}_i - \share{update}_i$;
\ENDFOR
\STATE  $P_i$ obtains $\share{\theta_T}_i$;
\end{algorithmic}
\end{algorithm}

\vspace{-1mm}
\section{optimization methods for the training process}~\label{efficiency_optimization}
In this section, we introduce two optimization methods on the accuracy and efficiency of MPL. The core ideas of the optimization methods are to simplify the structure of trained models and finish parts of computations locally, thus reducing the random noise added in the training process and the secure computations required for the training process.

\vspace{-1mm}
\subsection{Data-independent Feature Extraction Method}\label{subsec:separate}
\noindent\textbf{Reason to Simplify the Trained Model in MPL.} 
From the perspective of DP, the complex structures of deep neural network models require adding much random noise during the training process, thus resulting in large accuracy loss. First, the gradient vectors of deep neural network models are mainly high-dimension~\cite{dimension_reduction}. In DPSGD, as random noise is added to each gradient vector component, gradient vectors with higher dimensions require more noise. Bassily et al.~\cite{bassily2014private} theoretically prove that DPSGD implies extra loss with lower bound linear in the dimension of the gradient vectors. Second, deep neural network models have many redundant parameters. Thus adding noise to these redundant parameters would cause extra accuracy loss to neural network models. The empirical results of Section~\ref{subsec:end-to-end} further verify the above claim. Yu et al.~\cite{dimension_reduction} show that although optimizing original DPSGD with the assistance of a little public data, the accuracy gap between differentially private and non-private deep neural network models still keeps more than 15\% for CIFAR-10.

On the other hand, from the perspective of SMPC, the representation power of deep neural network models comes at the cost of efficiency, i.e. deep neural network models have many redundant parameters to fit the data. Learning these parameters requires a large number of computations. For example, training a ResNet-50 model~\cite{7780459}, a popular deep neural network model, on the ImageNet-1k dataset requires about $10^{18}$ single-precision operations~\cite{10.1145/3225058.3225069}. In the centralized setting, domain-specific processors (e.g. GPU) and distributed computation techniques can be used to accelerate the training process. However, when training a deep neural network model with MPL, the huge communication overhead brought by SMPC protocols would become the bottleneck of performance.

In linear secret sharing protocols, to complete one fixed-point number multiplication, it is necessary for parties to exchange messages. Therefore, training a deep neural network model on high-dimension data inevitably implies a huge communication overhead. For example, \texttt{Falcon}~\cite{wagh2020falcon}, one communication-efficient MPL framework, requires about 46 TB data transfer and more than one month to train a VGG-16 model~\cite{vgg16} on CIFAR-10 for 25 epochs in the LAN setting. \texttt{CryptGPU}, which utilizes GPU to accelerate the computation, still requires ten days to train the same model with \texttt{Falcon}~\cite{cryptGPU, wagh2020falcon}. 
In summary, with the current secret sharing protocols and differentially private optimization algorithms, it is still impractical to securely train an accurate and differentially private end-to-end deep neural network model.  

\noindent\textbf{Method.} In order to securely train a differentially private model with small accuracy loss and high efficiency simultaneously, we let each party first apply data-independent feature extractors, including public foundation models and heuristic feature extraction rules, to extract high-level features of input data.  After that, all parties train a shallow model (e.g. linear models or shallow neural network models) on the extracted features. In recent years, training models based on foundation models~\cite{foundation_model,jumper2021highly, rothchild2021c5t5,chen2021decision}, which are deep neural network models pre-trained on large-scale public data, has rapidly emerged as a new paradigm of the artificial intelligence field.  By utilizing the knowledge transferred from large-scale public data, foundation models can effectively extract high-level features of input data from various domains. Thus the models trained on the extracted features can be very accurate. Classical feature extraction rules also have proven their effectiveness on various tasks~\cite{kobayashi2013bfo,dang2016quality}. 
We then analyze how the above method reduces the accuracy loss brought by DP and improves the efficiency of MPL. Because foundation models and feature extraction rules are both data-independent, parties only need to add noise in the shallow model training process. Recent studies~\cite{iyengar2019towards, tramer2021differentially} have shown that we can apply DP to train shallow models with small accuracy loss privately. Therefore, training models on extracted features can significantly reduce the accuracy loss brought by DP. We further empirically verify this claim in Section~\ref{experiment}. In addition, as feature extraction is performed locally on plaintext data, parties only need to perform the shallow model training over data in the secret shared form. Thus the number of secure computations can be significantly reduced. In summary, extracting features with data-independent feature extractors enables parties to reduce the large accuracy loss and huge communication overhead of securely training a differentially private end-to-end deep neural network model.

\noindent\textbf{DP Guarantee of Data-independent Feature Extraction Method.} Because we perform the global model training process on secret shared data, the extracted features are invisible to potential adversaries (i.e. other parties, the users of trained models). Meanwhile, as we utilize secure DPSGD to protect the trained model, the feature extraction operation is oblivious to the adversaries.  In other words, \wqruan{the data-independent feature extraction can be viewed as a data-independent data pre-processing step, such as adding 1 to the features of data samples. Therefore, it should be unnecessary to add random noise to the data-independent feature extraction process.} This conclusion is the same as those of previous studies~\cite{yu2022differentially, tramer2021differentially}. 


\vspace{-1mm}
\subsection{Local Data-based Global Model Initialization Method }\label{subsec:initialization}
Utilizing the local data held by each party to initialize an accurate global model would significantly reduce the number of iterations in the global model training process, thus improving the efficiency of MPL. Hereby, we introduce how to aggregate local models from all parties to initialize the global model.

\noindent \textbf{Aggregation Methods.} Referring to the classical aggregation method in the Federated Learning field~\cite{yang2019federated} and considering that training data might be not independent and identically distributed (Non-IID for short) among parties, we propose two methods to aggregate the local models:
\begin{itemize}
    \item \textbf{Averaging Strategy.} Averaging local model parameters as parameters of the initial global model. This method is originated from the literature of federate learning~\cite{yang2019federated}. 
        
    \item \textbf{Accuracy Strategy.} When data are Non-IID, initializing the global model by averaging local models might cause that the accuracy of the initial global model is lower than local models. Therefore, an alternative strategy is to choose the most accurate local model as the initial model.
\end{itemize}

With the above aggregation methods, we show the global model initialization method in Protocol~\ref{alg:global_model_initialization}. Because the distribution status of data is difficult to measure directly, we choose an initial global model by selecting the most accurate candidate model generated by different aggregation methods. We show the details of the initial global model selection method in Lines 4-11 of Protocol~\ref{alg:global_model_initialization}. Parties generate candidate models with different aggregation methods and evaluate the accuracy of each candidate model, respectively. Finally, all parties obtain the most accurate initial global model by comparing the accuracy of candidate models.  We analyze the security and communication complexity of Protocol~\ref{alg:global_model_initialization} in  Appendix~\ref{sec:analysis_initialization},  where we introduce the details of the two aggregation methods.


\noindent \textbf{Privacy Guarantee.} 
Benefit from the post-processing immunity property of DP~\cite{foundation_dp}, we only need to analyze the privacy guarantee of the local model training phase. In the following theorem, we show the privacy guarantee of Protocol~\ref{alg:global_model_initialization} respect to the logically global dataset $D = \{D_1, D_2, \cdots, D_m\}$ (the proof of Theorem~\ref{theo:model_initialization} is shown in Appendix~\ref{Missing Proof}).

\begin{theorem}\label{theo:model_initialization}
Protocol~\ref{alg:global_model_initialization} satisfies ($\epsilon_1, \delta_1$)-DP.
\end{theorem}
\begin{algorithm}
\small
\caption{Local Data-based Global Model Initialization }
\label{alg:global_model_initialization}
\begin{algorithmic}[1]
\REQUIRE  $P_i$ holds a local dataset $D_i$, the privacy parameters $\epsilon_1$ and $\delta_1$, clipping bound $C_1$ and aggregation method set $A = \{A_1, A_2\}$;
\ENSURE $P_i$ obtains the share of initial global model $\share{\theta_{c}}_i$;
\STATE $P_i$ trains a local model $\theta^{l}_{i}$ with Protocol~\ref{DPSGD} and $\epsilon = \epsilon_1, \delta = \delta_1, C = C_1$;
\STATE $P_i$ sends $\share{\theta^{l}_{i}}_{j \neq i}$ to $P_{j \neq i}$;
\STATE Parties randomly initialize a candidate model $\theta_{c}$ and securely evaluate the accuracy of $\theta_{c}$ as $Accuracy_c$;
\FOR{Each $A_k \in A$}
\STATE $P_i$ obtains the share of kth candidate model $\share{\theta_{ck}}_i = A_k(\share{\theta^{l}_{1}}_i, \share{\theta^{l}_{2}}_i,\cdots, \share{\theta^{l}_{m}}_i)$;
\STATE All parties collaboratively evaluate the accuracy of $\theta_{ck}$ as $Accuracy_{ck}$;
\IF{$Accuracy_{ck} \geq$  $Accuracy_c$}
\STATE $\share{\theta_{c}}_i = \share{\theta_{ck}}_i$;
\STATE $Accuracy_c = Accuracy_{ck}$;
\ENDIF
\ENDFOR
\STATE $P_i$ obtains $\share{\theta_{c}}_i$ as the share of the initial global model;
\end{algorithmic}
\end{algorithm}

\vspace{-1mm}
\subsection{Put Things Together: \wqruan{PEA}}~\label{subsec:summary}
Combining the secure DPSGD protocol and the above two optimization methods, we come to \wqruan{PEA} shown in Protocol~\ref{alg:private_efficient_MPL}. Each party first extracts features of his or her input dataset with data-independent feature extractors. Note that this step is completed locally. And the following training processes are performed on the extracted features. Next, they collaboratively initialize the global model through using the method introduced in Protocol~\ref{alg:global_model_initialization}. Finally, all parties improve the global model on the logically global dataset with the secure DPSGD protocol introduced in Section~\ref{secure DPSGD}. As the privacy guarantees of Protocol~\ref{alg:secure_dp_sgd} and Protocol~\ref{alg:global_model_initialization} have been shown in Theorem~\ref{theo:secure_dp_sgd} and Theorem~\ref{theo:model_initialization}, with the sequential composition theorem introduced in Section~\ref{sec:differential_privacy}, we can show that Protocol~\ref{alg:private_efficient_MPL} satisfies ($\epsilon_1+\epsilon_2, \delta_1 + \delta_2$)-DP.
\begin{theorem}\label{main_privacy_theorem}
Protocol~\ref{alg:private_efficient_MPL} satisfies ($\epsilon_1+\epsilon_2, \delta_1 + \delta_2$)-DP.
\end{theorem}

\begin{algorithm}
\caption{\wqruan{Private, Efficient, and Accurate Training of MPL}}
\label{alg:private_efficient_MPL}
\small
\begin{algorithmic}[1]
\REQUIRE  $P_i$ holds a local dataset $D_i$, the privacy parameters $\epsilon_1, \epsilon_2$ and $\delta_1, \delta_2$, clipping bound $C_1, C_2$.
\ENSURE $P_i$ obtains the share of trained global model $\share{\theta_{T}}_i$.
\STATE $P_i$ extracts features of the local dataset $D_i$ as $FD_i$ with data-independent feature extractors;
\STATE All parties apply Protocol~\ref{alg:global_model_initialization} to initialize the global model with privacy parameter ($\epsilon_1, \delta_1$) and clipping bound $C_1$;
\STATE All parties send the shares of local datasets to other parties;
\STATE All parties obtain the trained global model through applying secure DPSGD protocol described in Protocol~\ref{alg:secure_dp_sgd} to improve the initial global model with privacy parameters ($\epsilon_2, \delta_2$) and clipping bound $C_2$;
\end{algorithmic}
\end{algorithm}

\vspace{-2mm}
\section{Implementation and Evaluation}\label{experiment}
In this section, we first conduct an end-to-end comparison to illustrate the effectiveness of \wqruan{PEA}. Subsequently, we evaluate the running time and approximation error of the inverse of square root protocol, which is the main component of the secure DPSGD protocol. After that, we evaluate the effectiveness of optimization methods and compare the performance of different aggregation methods. Subsequently, we compare \wqruan{PEA} with two state-of-the-art methods on federated learning and secure aggregation with DP. Finally, we evaluate the accuracy of models trained by MPL to show that MPL can significantly improve the accuracy of differentially private models.

\vspace{-1mm}
\subsection{Implementation and Experiment Setup}~\label{subsec:experiment_setup}
\noindent\textbf{Implementation.} We implement \wqruan{PEA} in two open-source MPL frameworks: \texttt{TF-Encrypted}~\cite{TF-Encrypted}\footnote{\url{https://github.com/tf-encrypted/tf-encrypted}} and \texttt{Queqiao}\footnote{\url{https://github.com/FudanMPL/SecMML}}.

\texttt{TF-Encrypted} is implemented in Python and contains multiple back-end SMPC protocols. We implement \wqruan{PEA} based on its supported \texttt{ABY3}~\cite{mohassel2018aby3} back-end protocol. As \texttt{ABY3} is designed on the integer ring rather than the prime field, we implement some primitives involved in Protocol~\ref{alg:secure_inversesqrt} in \tfencrypted by combining some primitives of \texttt{ABY3} with other protocols. For example, we implement $\mathsf{Bit\_Dec}$ through combining $\mathsf{A2B}$ protocol and the sharing conversion protocol proposed by Knott et al.~\cite{crypten2020}. We set the ring size as $2^{64}$ and the bit number of the fractional part of the fixed-point number $f$ as 20 in \tfencrypted.

\texttt{Queqiao} is implemented in C++ based on BGW protocol~\cite{ref_bgw}. It supports secure model training on the distributed data from three+ parties. As some primitives involved in Protocol~\ref{alg:secure_inversesqrt} are not implemented by \texttt{Queqiao}, we supplement the missed primitives in \texttt{Queqiao} to implement \queqiao. Finally, we set the field size as $10 ^ {17} + 3$ and $f$ as 20 in \queqiao.

\noindent\textbf{Datasets.}  We introduce three widely-used datasets involved in our experiments as follows: 
\begin{itemize}
    \item\textbf{MNIST~\cite{MNIST}} is a handwritten digit dataset, which is widely used in the evaluation of multi-party learning studies~\cite{mohassel2018aby3,cryptGPU}. MNIST contains 60,000 training samples and 10,000 testing samples. Each data sample represents one digit between 0 to 9 and has 28 $\times$ 28 gray pixels.
    \item \textbf{CIFAR-10~\cite{CIFAR-10}} contains 60,000 32 $\times$ 32 RGB images in ten classes, and each class has 6,000 images. We thus randomly sample 50,000 images as the training dataset and let the rest of 10,000 images as the testing dataset.
    \item \textbf{IMDb~\cite{IMDb}} is a popular dataset in the natural language processing field. It contains 50,000 film reviews from IMDb with 25,000 training samples and 25,000 testing samples. The binary label of IMDb indicates the sentiment of film reviews (i.e. positive or negative).
\end{itemize}

\noindent\textbf{Dataset Partition.} We partition each dataset into local datasets by randomly dividing it into $m$ disjointed parts, where $m$ is the number of parties. In experiments of Sections~\ref{subsec:end-to-end}, \ref{subsec:e_a_inverse},  \ref{subsec:effectiveness_optimizations}, \ref{subsec:compare_aggregation},and \ref{subsec:FL_secure_DP}, we set $m$ as 3, which is the party number \tfencrypted and \queqiao both support.

\noindent\textbf{Feature Extractors and Classification Models.} We employ HOG~\cite{dalal2005histograms} and two foundation models in the computer vision and natural language processing (NLP) fields, namely \wqruan{SimCLR}~\cite{simclr} and BERT~\cite{BERT}, to extract features of MNIST, CIFAR-10, and IMDb, respectively. Moreover, in the following experiments, we use logistic regression as the classification model trained on the extracted features.

\noindent\textbf{Experiment Settings.} We run experiments on three Linux servers, each of them with 2.7 GHZ of CPU and 128GB of RAM. We consider two types of network environments. The first one is the LAN setting with 1Gbps bandwidth and negligible latency. The second one is the WAN setting with 100 Mbps bandwidth and 20ms round-trip-time latency. We simulate these two network environments through using the tc tool\footnote{\url{https://man7.org/linux/man-pages/man8/tc.8.html}}. To avoid the bias brought by the randomness of noise sampling, we run all experiments that involve randomness five times and plot the corresponding error bars in figures.

\noindent\textbf{Parameter Settings.} We set the batch size $B$ as 128 and the gradient clipping bound $C$ as 3. In addition, we set $\delta$ as $\frac{1}{10n}$, where $n$ is the size of the training dataset. For the privacy budget $\epsilon$, we set it as 2 and run all global model training processes for two epochs. We use the privacy budget accountant tools\footnote{\url{https://github.com/tensorflow/privacy/blob/7eea74a6a1cf15e2d2bd890722400edd0e470db8/research/hyperparameters_2022/rdp_accountant.py}} released by Google to set the parameters of Gaussian distributions. 

\vspace{-1mm}
\subsection{End-to-end Comparison}\label{subsec:end-to-end}

\noindent\textbf{Baselines.} We employ several popular differentially private deep neural network models trained on raw datasets as baselines. For MNIST and CIFAR-10, we train the differentially private LeNet model~\cite{MNIST} and VGG-16 model~\cite{vgg16} respectively. The non-private training of these two models is also supported by \texttt{CryptGPU}~\cite{cryptGPU} and \texttt{Falcon}~\cite{wagh2020falcon}. For IMDb, we train the same differentially private neural network model that is used by Bu et al.~\cite{Bu2020Deep} as the baseline. We refer to it as SampleNet.  We set $\delta$ and the noise multiplier of Gaussian distribution as the same value as ours during the baseline training. Then we tune $\epsilon$ as the iteration number increases. All baseline models are trained by \texttt{Opacus} framework released by Yousefpour et al.~\cite{DBLP:journals/corr/abs-2109-12298}. We also apply a grid search procedure to find the best hyperparameter configuration for them. The potential hyperparameters are listed in Table~\ref{tab:baseline_grid_search_paramters} of Appendix~\ref{parameter setting}. 

\noindent\textbf{Privacy Budget Allocation.} With local datasets, we train local classification models with DPSGD.  In this phase, we set $\epsilon_1 = 0.25$ and $\delta_1 = \frac{1}{100n}$. We show the values of other hyperparamters in Table~\ref{tab:local_model_paramters} of Appendix~\ref{parameter setting}. For the global model training, we set $\delta_2$ as $\frac{9}{100n}$ and tune the total privacy budget $\epsilon$ as the iteration number increases.

\begin{figure*}[ht]
    \centering
    \subcaptionbox{\tfencrypted\_MNIST
    \label{fig:e2e_mnist_tf}}[0.28\linewidth]
    {
        \includegraphics[width=\linewidth]{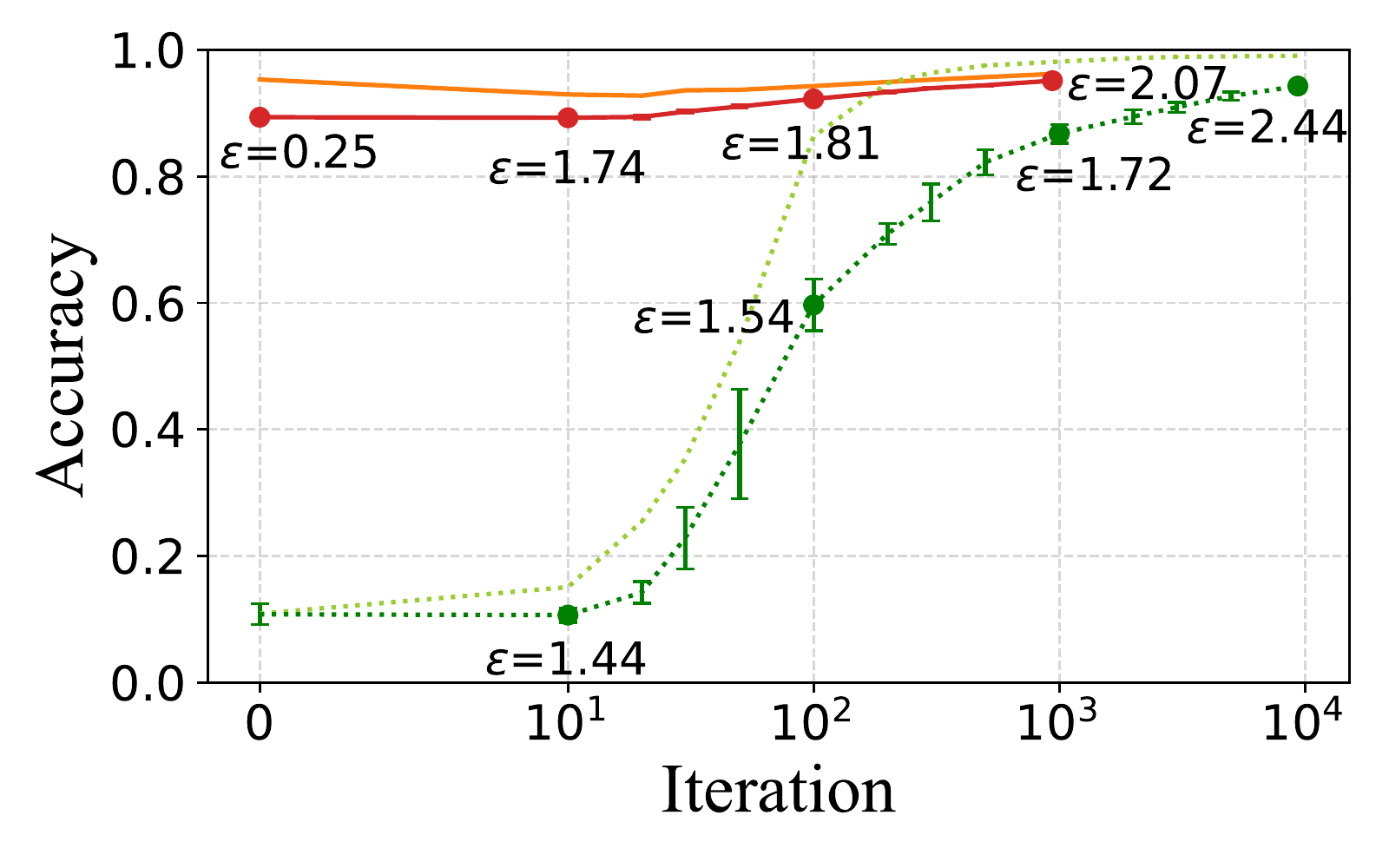}
    }\quad%
    \subcaptionbox{\tfencrypted\_CIFAR-10
    \label{fig:e2e_cifar_tf}}[0.28\linewidth]
    {
        \includegraphics[width=\linewidth]{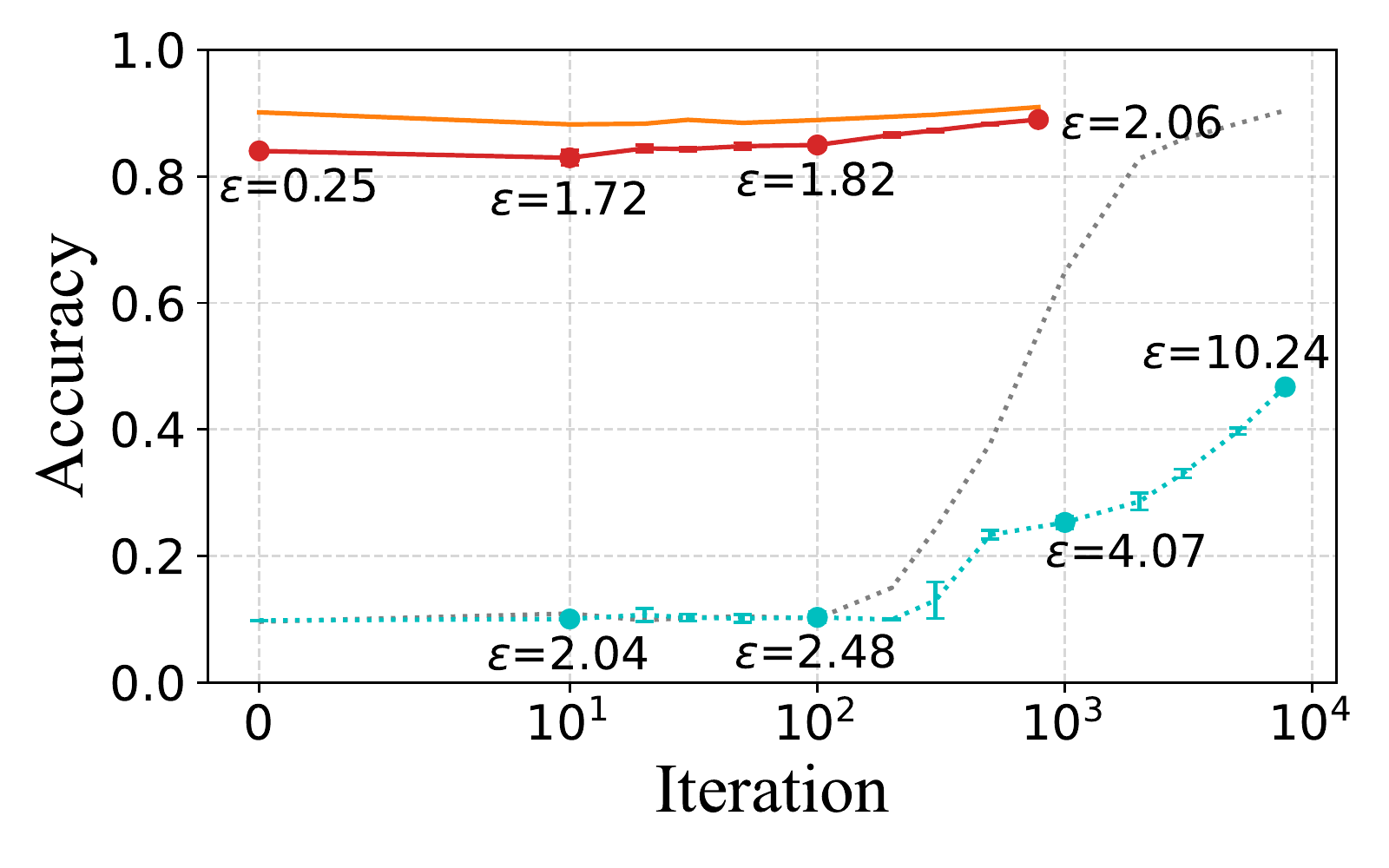}
    }\quad%
    \subcaptionbox{\tfencrypted\_IMDb
    \label{fig:e2e_imdb_tf}}[0.28\linewidth]
    {
        \includegraphics[width=\linewidth]{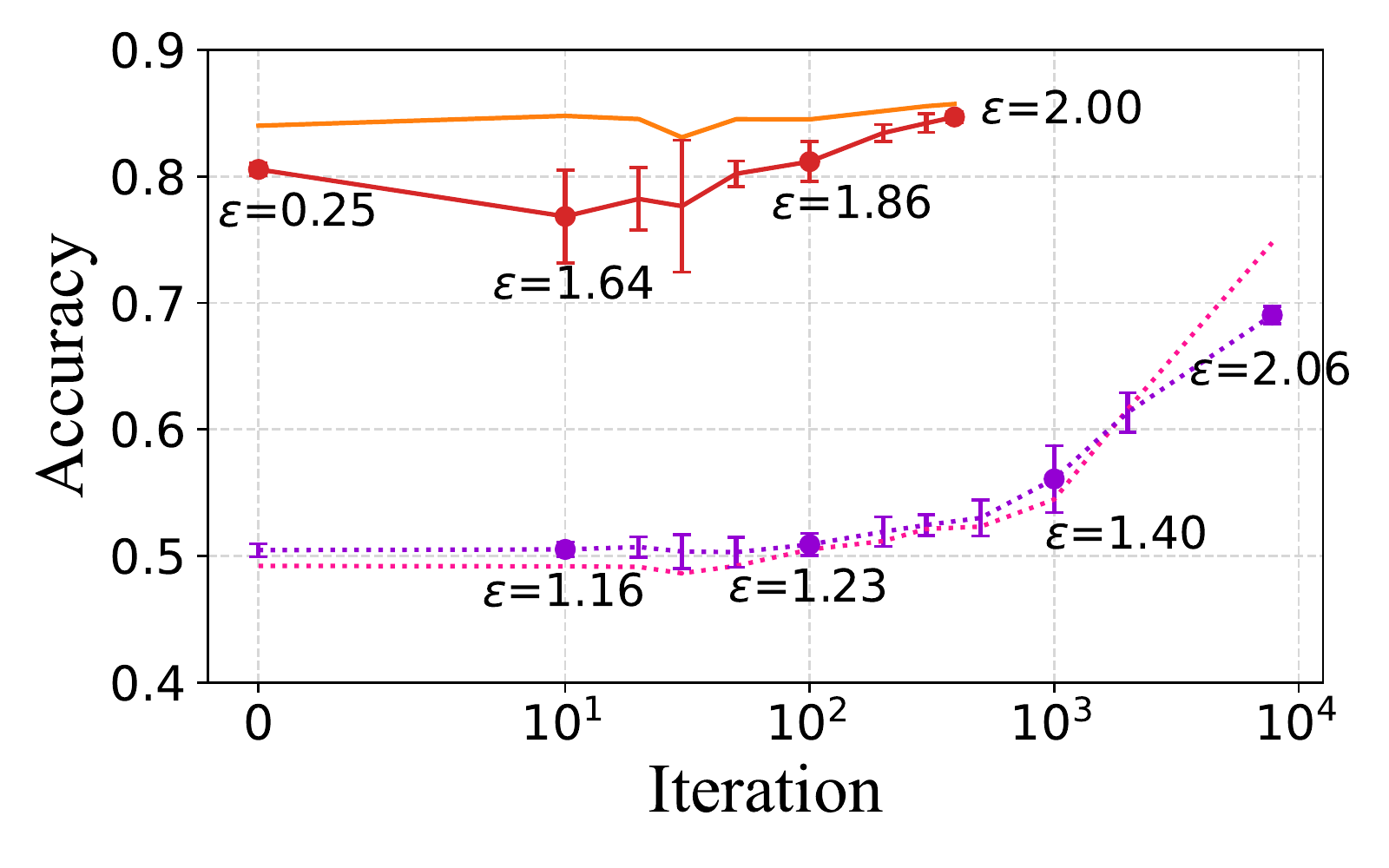}
    }
        \subcaptionbox{\queqiao\_MNIST
    \label{fig:e2e_mnist_qeuqiao}}[0.28\linewidth]
    {
        \includegraphics[width=\linewidth]{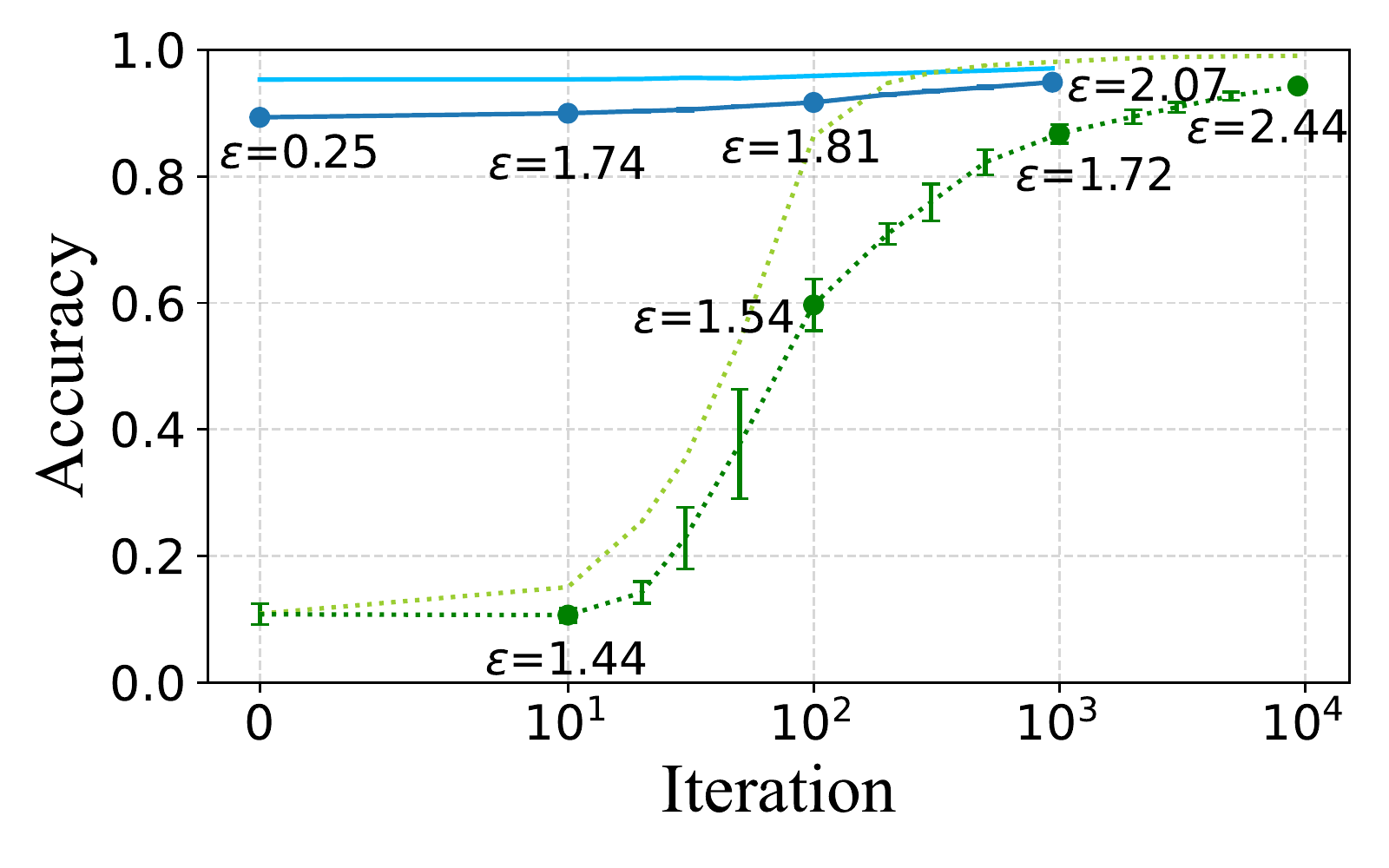}
    }\quad%
    \subcaptionbox{\queqiao\_CIFAR-10
    \label{fig:e2e_cifar_queqiao}}[0.28\linewidth]
    {
        \includegraphics[width=\linewidth]{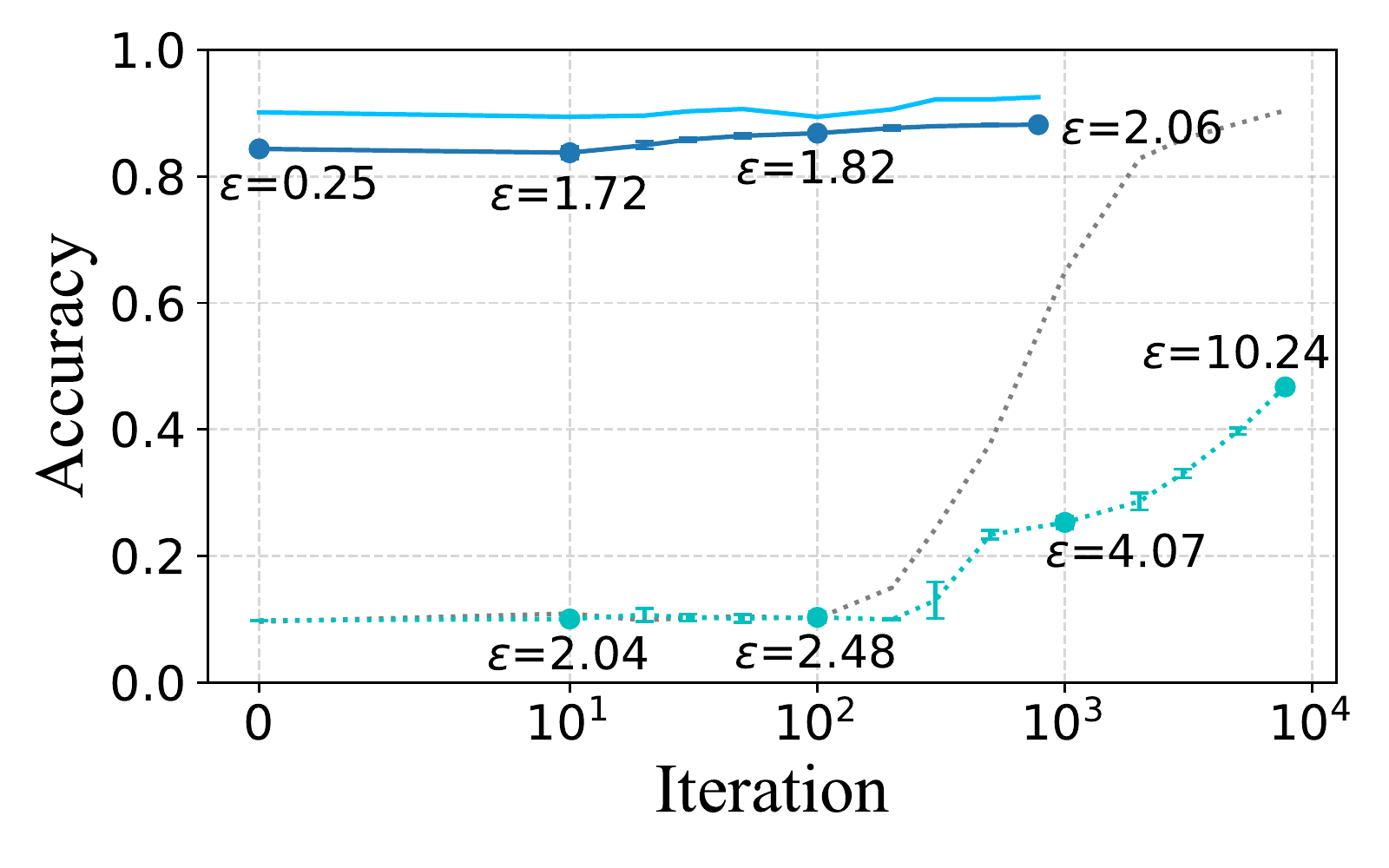}
    }\quad%
    \subcaptionbox{\queqiao\_IMDb
    \label{fig:e2e_imdb_queqiao}}[0.28\linewidth]
    {
        \includegraphics[width=\linewidth]{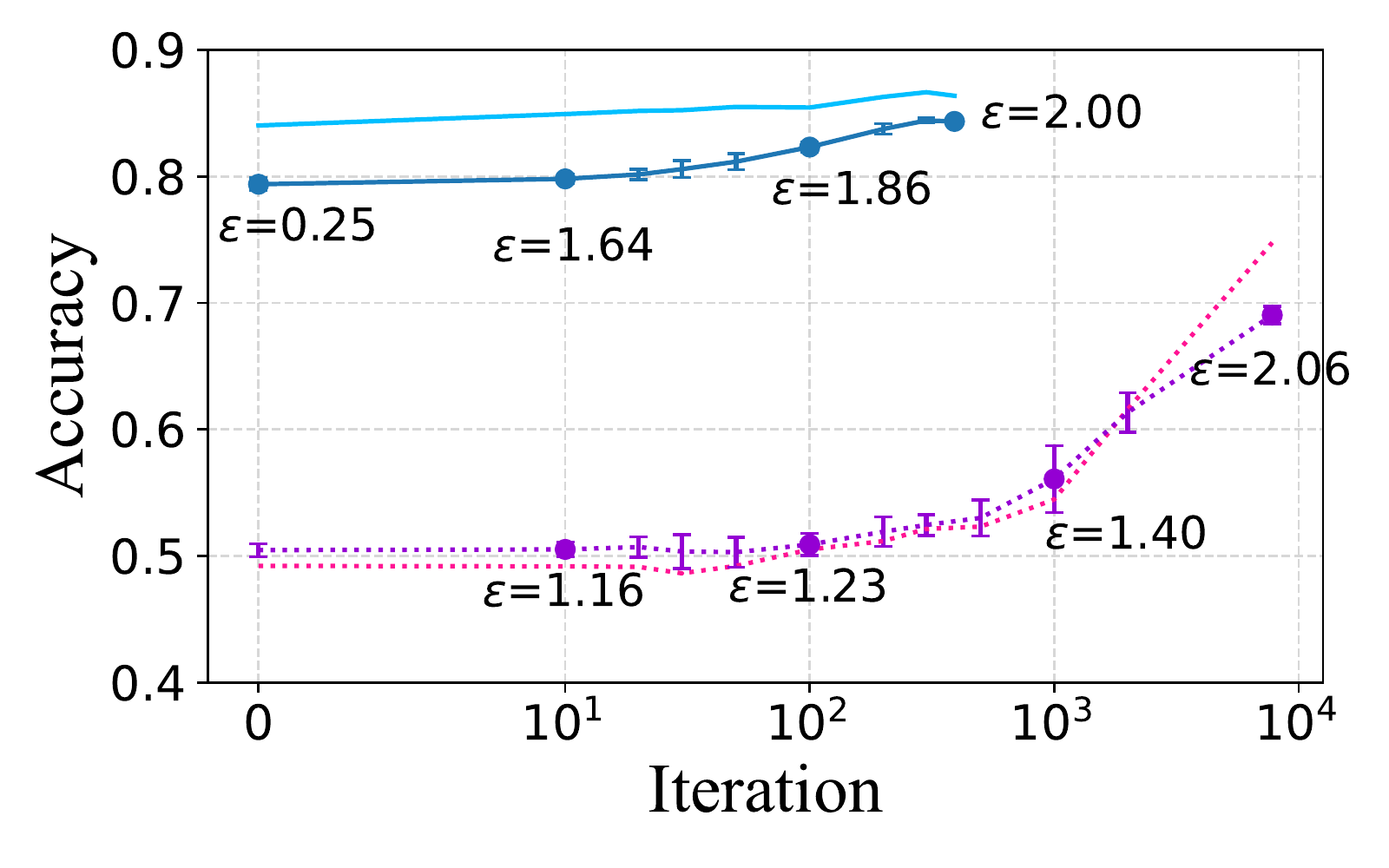}
    }
    \includegraphics[width=0.8\linewidth]{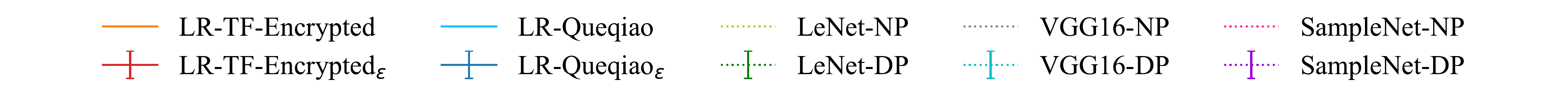}
    
    \caption{End-to-end comparison of different model training methods. LR-TF-Encrypted$_\epsilon$ and LR-Queqiao$_\epsilon$ are our proposed methods. LR-TF-Encrypted and LR-Queqiao are non-private logistic regression models trained by \texttt{TF-Encrypted} and \texttt{Queqiao}. 
    }
    \label{fig:e2e_comparision}
\end{figure*}

\noindent\textbf{Results.} As is shown in Figure~\ref{fig:e2e_comparision}, under the same number of iterations, the differentially private logistic regression models trained by \tfencrypted and \queqiao have higher accuracy than differentially private deep neural network models in all three datasets. Note that the baseline models trained on different datasets have different privacy budgets as we set different batch sizes for them. The detailed parameter settings are shown in Table~\ref{tab:baseline_paramters} of Appendix~\ref{parameter setting}.

As to the efficiency, since we only need to train a shallow model, the time consumption of one iteration is much smaller than that of the deep neural network model training. For example, the training of differentially private classifiers for CIFAR-10 costs about 0.5 seconds and 4 seconds per iteration in \tfencrypted and \queqiao, while \texttt{CryptGPU} requires about 12 seconds to complete one iteration of a non-private VGG-16 model training, which serves as the baseline. As a result, \tfencrypted and \queqiao can train a differentially private classifier for CIFAR-10 with an accuracy of more than 88\% within 7 minutes and 55 minutes, while \texttt{CryptGPU} requires more than 16 hours to train a non-private model to achieve the same accuracy. 

In addition, as we only add noise to the linear model training process, the accuracy gaps between differentially private and non-private linear models are all smaller than those of differentially private and non-private deep neural network models. These results further verify the claim of Section~\ref{subsec:separate}.

\vspace{-1mm}
\subsection{Efficiency and Approximation Error of Protocol~\ref{alg:secure_inversesqrt}}\label{subsec:e_a_inverse}
\noindent\textbf{Baseline.} We compare the efficiency and accuracy of Protocol~\ref{alg:secure_inversesqrt} with \texttt{CrypTen}~\cite{crypten2020}, an MPL framework released by Facebook. As \texttt{CrypTen} cannot directly compute the inverse of square root, we implement it by sequentially calling the sqrt() and the reciprocal() APIs of \texttt{CrypTen}. Note that we run all three frameworks in the 3-party scenario and compare the online running time of all MPL frameworks here.


\begin{table}
\centering
\footnotesize
\caption{Online running time (in ms) comparison of the inverse square root protocols. We run all experiments ten times and report the average running time. The shortest running times are marked in bold.}
\scalebox{0.85}{
\begin{tabular}{|c|l|l|l|l|l|}
\hline
\multicolumn{1}{|c|}{\multirow{2}{*}{\textbf{Setting}}} & \multicolumn{1}{c|}{\multirow{2}{*}{\textbf{Framework}}} & \multicolumn{4}{c|}{\textbf{Batch size}} \\ \cline{3-6} 
\multicolumn{1}{|c|}{}                         & \multicolumn{1}{c|}{}                            & \multicolumn{1}{c|}{1}     & \multicolumn{1}{c|}{32}    & \multicolumn{1}{c|}{64}   & \multicolumn{1}{c|}{128}    \\ \hline
\multirow{3}{*}{LAN}                 
                                               & \tfencrypted                              & 28.22         &  \textbf{28.75}    & \textbf{29.28}    &   \textbf{29.93}   \\ \cline{2-6} 
                                                    & \queqiao                                     & \textbf{22.94}        & 76.43    & 123.37    &  233.52    \\ \cline{2-6} 
                                               & \texttt{CrypTen}                                          & 170.42            & 236.66      & 271.64      & 318.23  \\ \hline
\multirow{3}{*}{WAN}                           & \tfencrypted                             &  841.17         &  843.6     & 848.0   & 848.3    \\ \cline{2-6} 
                                               & \queqiao                                      & \textbf{456.8}           & \textbf{556.19}       & 
\textbf{595.79}    & \textbf{701.70}     \\ \cline{2-6} 
                                               & \texttt{CrypTen}                                          & 4696.42          & 6632.96   &    6641.89   &6647.64 \\ \hline
\end{tabular}}

\label{tab:efficiency_comparison}
\end{table}

\noindent\textbf{Efficiency of Protocol~\ref{alg:secure_inversesqrt}.} We show the running time comparison results in Table~\ref{tab:efficiency_comparison}. Under all batch sizes, \tfencrypted and \queqiao are more efficient than \texttt{CrypTen} in both LAN and WAN settings. This is because we use a polynomial to approximate the target function directly and only require constant communication rounds. In contrast, \texttt{CrypTen} uses the Newton-Raphson method to approximate the target function iteratively, thus requiring more communication overhead. 

In addition, under the LAN setting, the running time of \tfencrypted and \texttt{CrypTen} changes little as the batch size increases. The main reason is that these two frameworks are built on \texttt{TensorFlow} and \texttt{Pytorch}, which both support efficient vectorized computations. While in the WAN setting, where the communication round number is the dominant factor of efficiency, the running time of \queqiao is the smallest as it has the least communication round number. Note that we implement a part of primitives involved in Protocol~\ref{alg:secure_inversesqrt} in \tfencrypted by combining multiple protocols. Thus it has more communication rounds than \queqiao.

\begin{figure}[t]
    \centering
    \includegraphics[scale=0.35]{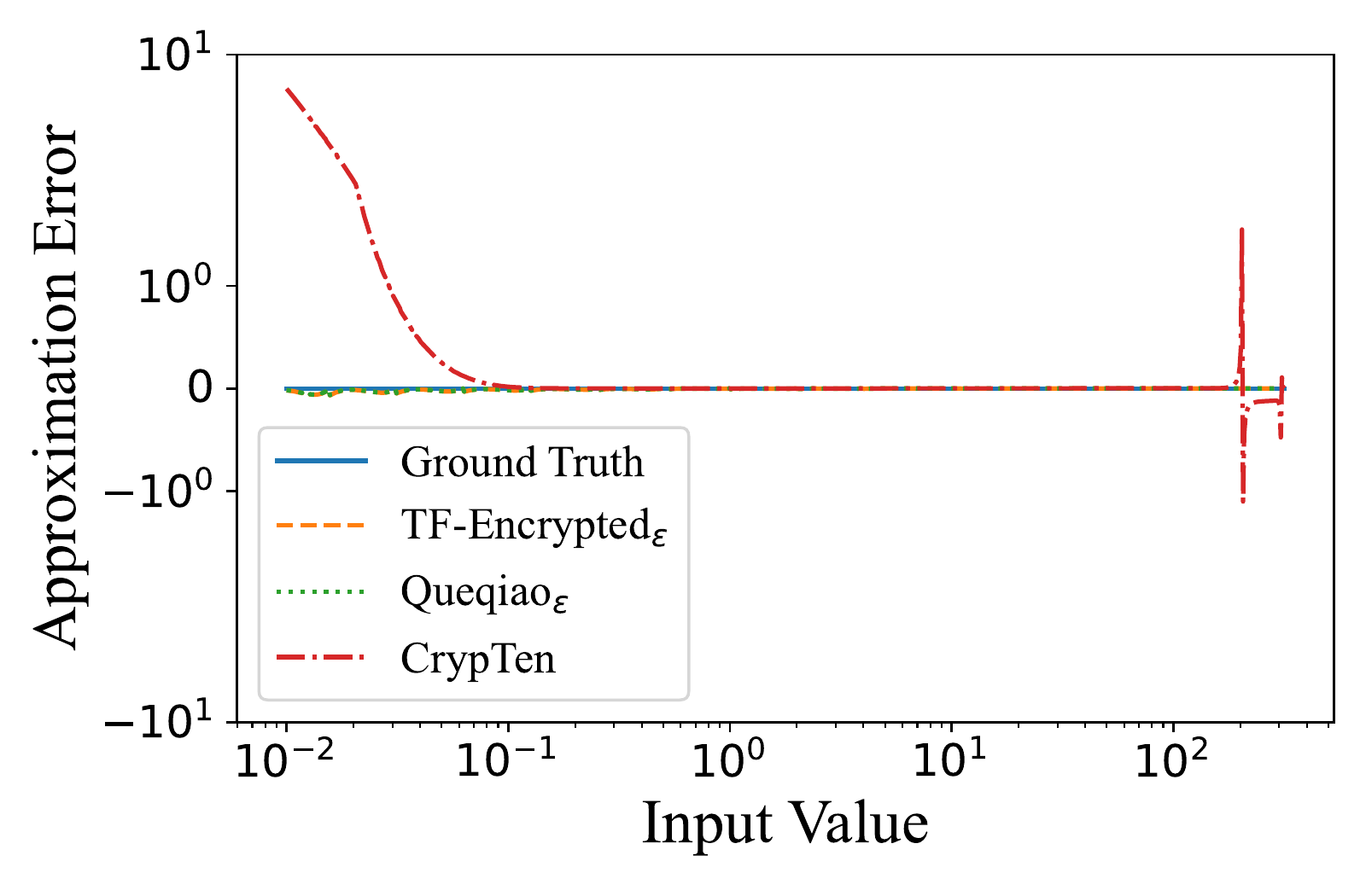}
    \caption{Approximation error comparison of the inverse square root computations. }
    \label{fig:bias}
\end{figure}

\noindent\textbf{Approximation Errors.} We show the approximation error, which is calculated through subtracting exact results by approximated results, of three frameworks in Figure~\ref{fig:bias}. Compared with \texttt{CrypTen}, which is accurate only when input values are higher than 0.1 and smaller than 200, the approximation errors of \tfencrypted and \queqiao stably keep small as the input value changes from 0.01 to 300. \texttt{CrypTen} has such a large error for parts of input values because it uses the Newton-Raphson method to approximate the target function. Its approximation errors are highly dependent on the pre-defined initial point and the iteration number. Therefore, \texttt{CrypTen} could be accurate for those input values in a fixed range while having a large error for other input values.  Besides, the approximation errors of \tfencrypted and \queqiao, which are  all smaller than 0, can prevent the privacy guarantee from being destroyed as we discuss in Section~\ref{Sec:Inverse}. 

\begin{figure*}[ht]
    \centering
    \subcaptionbox{MNIST
    \label{fig:rawdata_mnist}}[0.28\linewidth]
    {
        \includegraphics[width=\linewidth]{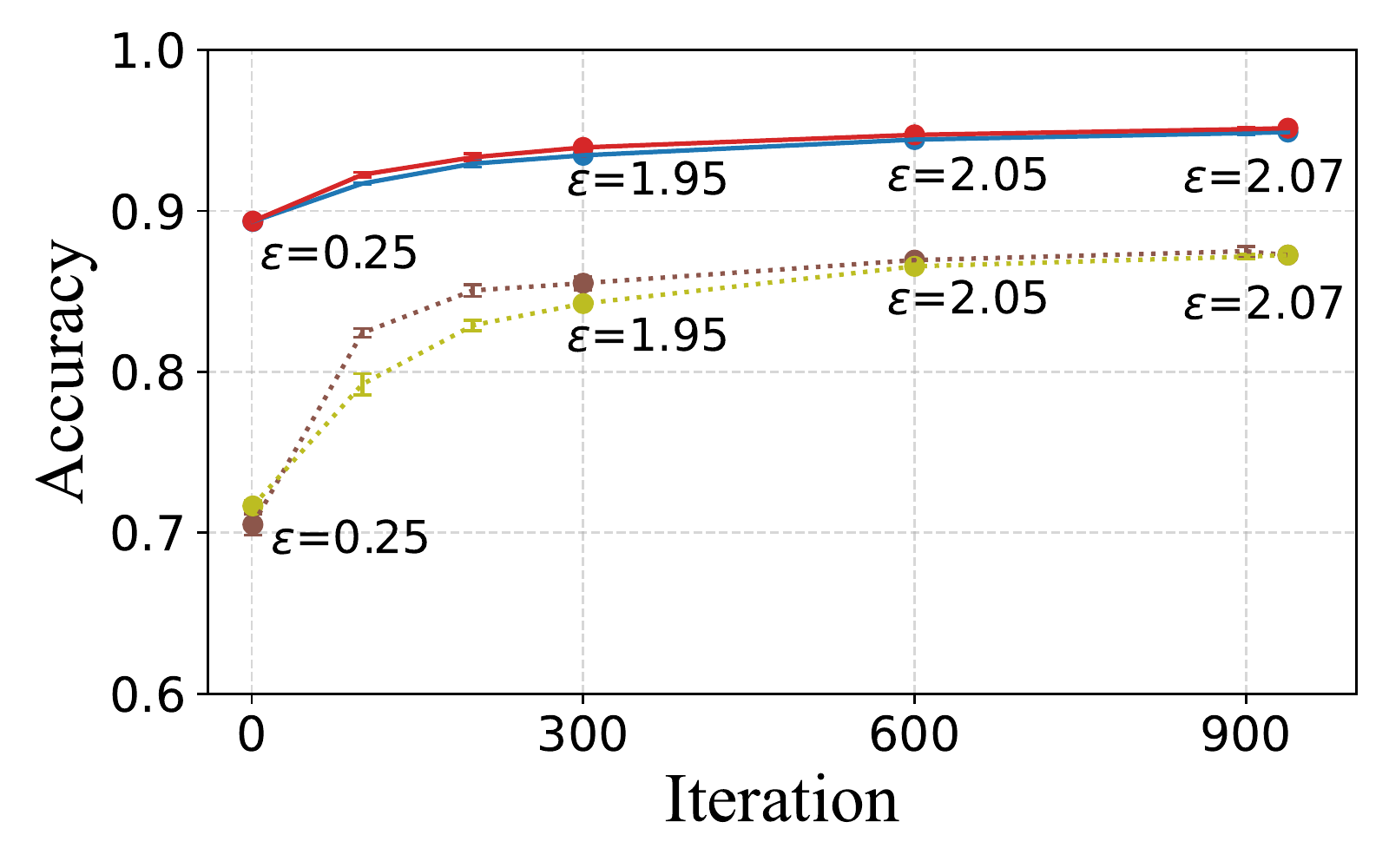}
    }\quad%
    \subcaptionbox{CIFAR-10
    \label{fig:rawdata_cifar}}[0.28\linewidth]
    {
        \includegraphics[width=\linewidth]{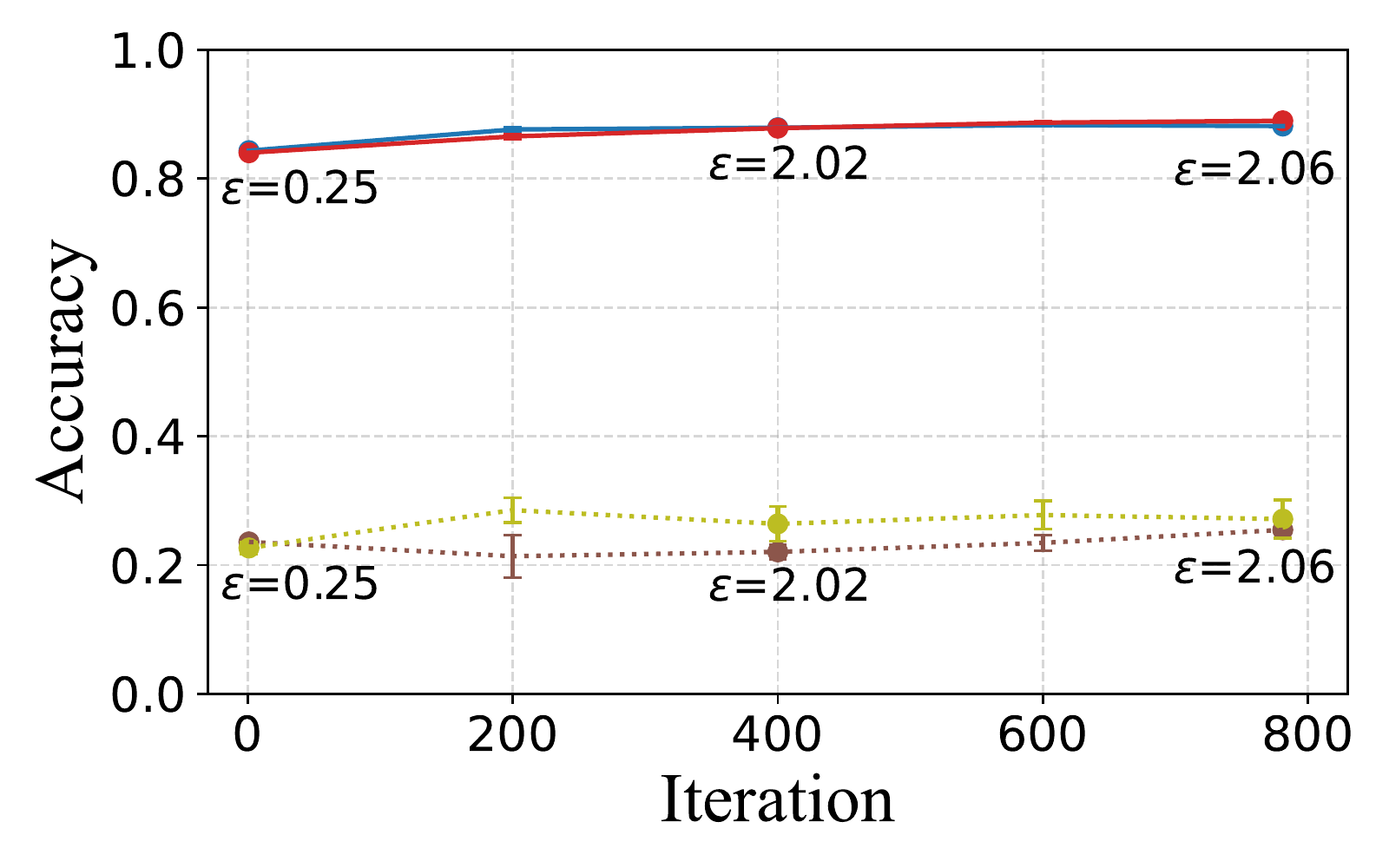}
    }\quad%
    \subcaptionbox{IMDb
    \label{fig:rawdata_imdb}}[0.28\linewidth]
    {
        \includegraphics[width=\linewidth]{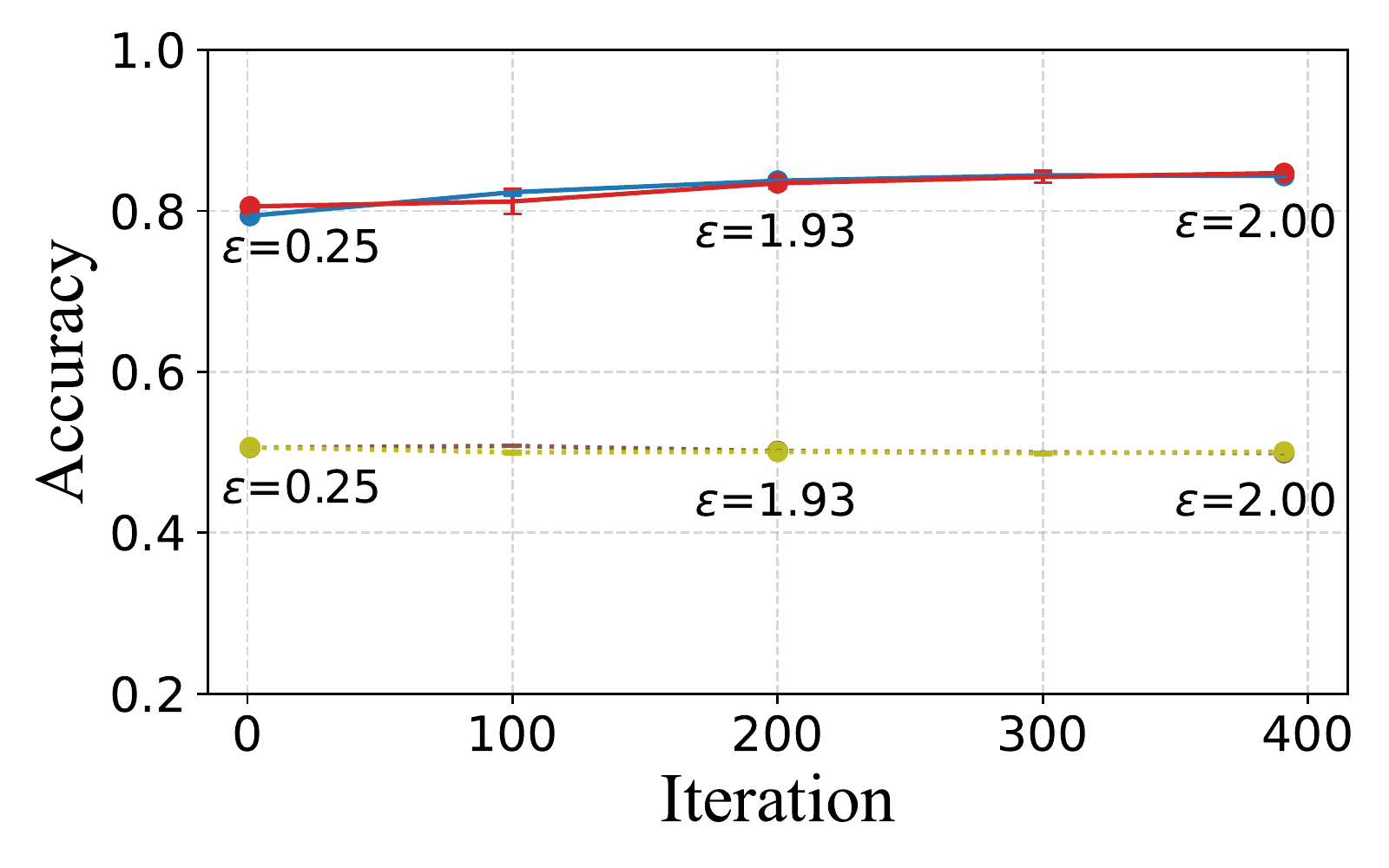}
    }
    \includegraphics[width=0.8\linewidth]{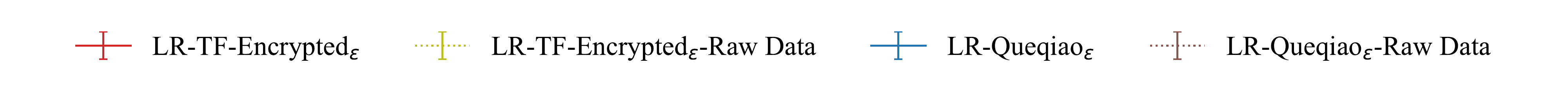}
    \caption{Accuracy comparison of the models trained on extracted features and raw data. LR-TF-Encrypted$_\epsilon$ and LR-Queqiao$_\epsilon$ are our proposed methods.}
    \label{fig:feature_extraction}
\end{figure*}

\begin{figure*}[ht]
    \centering
    \subcaptionbox{MNIST
    \label{fig:init_mnist}}[0.28\linewidth]
    {
        \includegraphics[width=\linewidth]{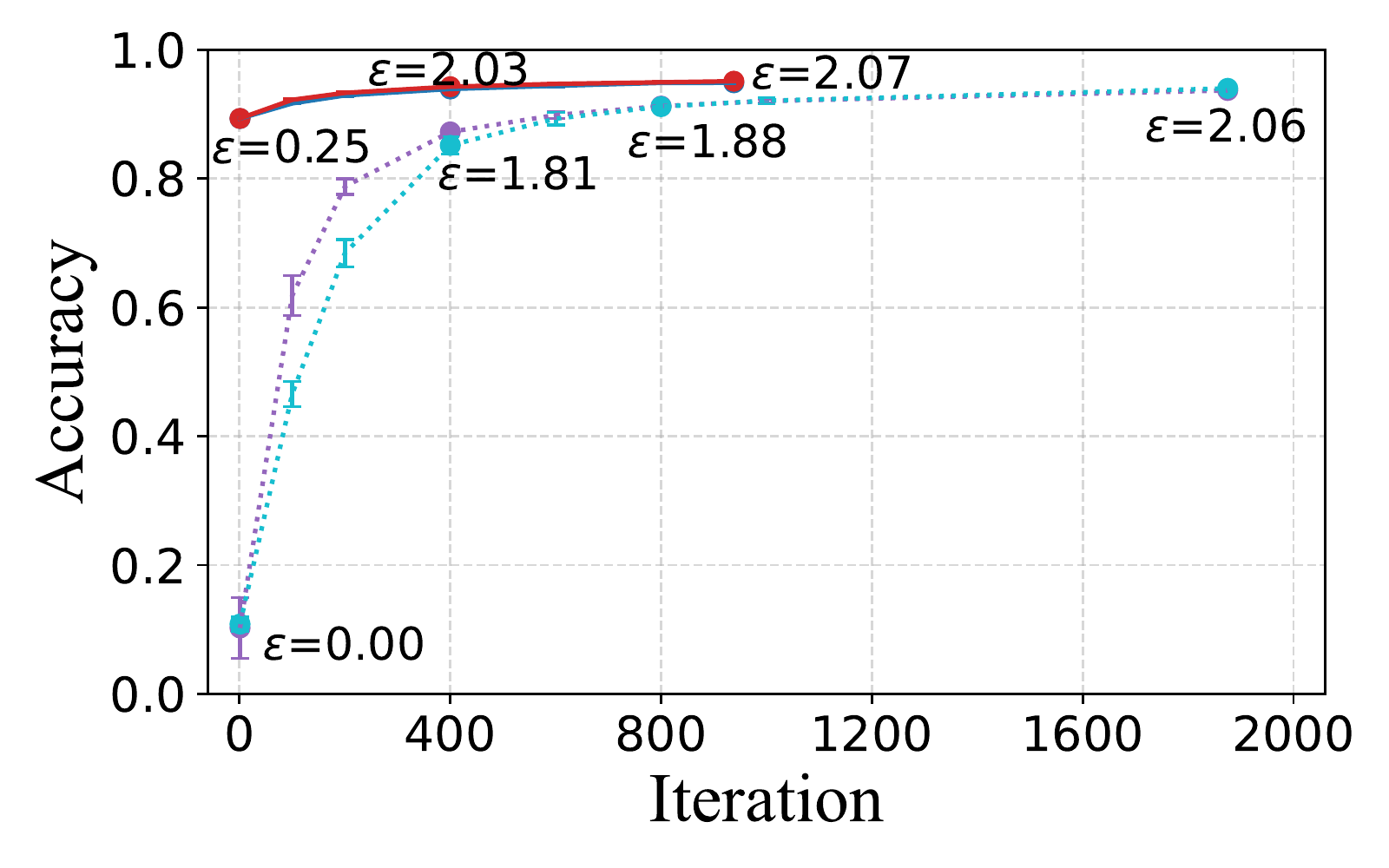}
    }\quad%
    \subcaptionbox{CIFAR-10
    \label{fig:init_cifar}}[0.28\linewidth]
    {
        \includegraphics[width=\linewidth]{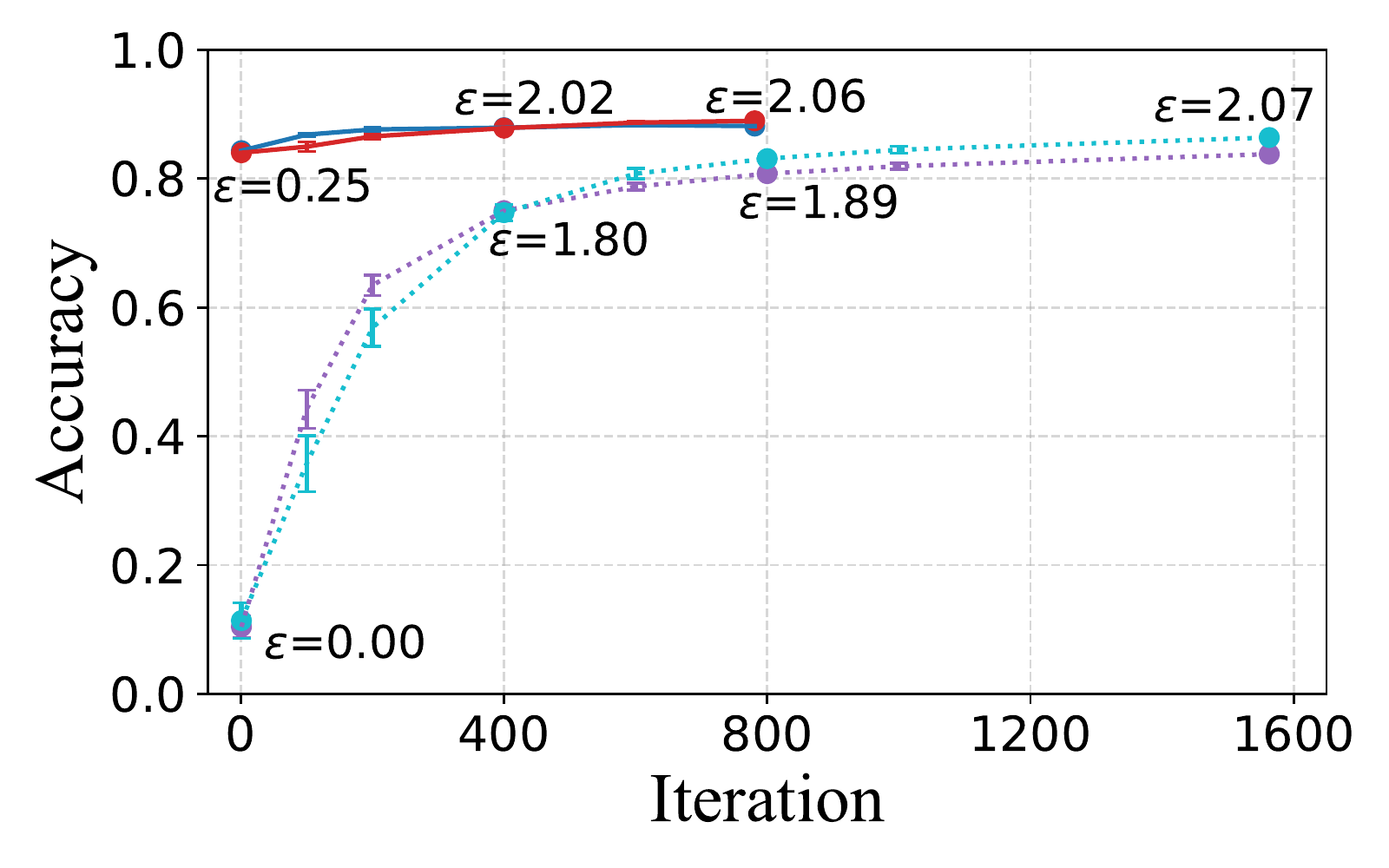}
    }\quad%
    \subcaptionbox{IMDb
    \label{fig:init_imdb}}[0.28\linewidth]
    {
        \includegraphics[width=\linewidth]{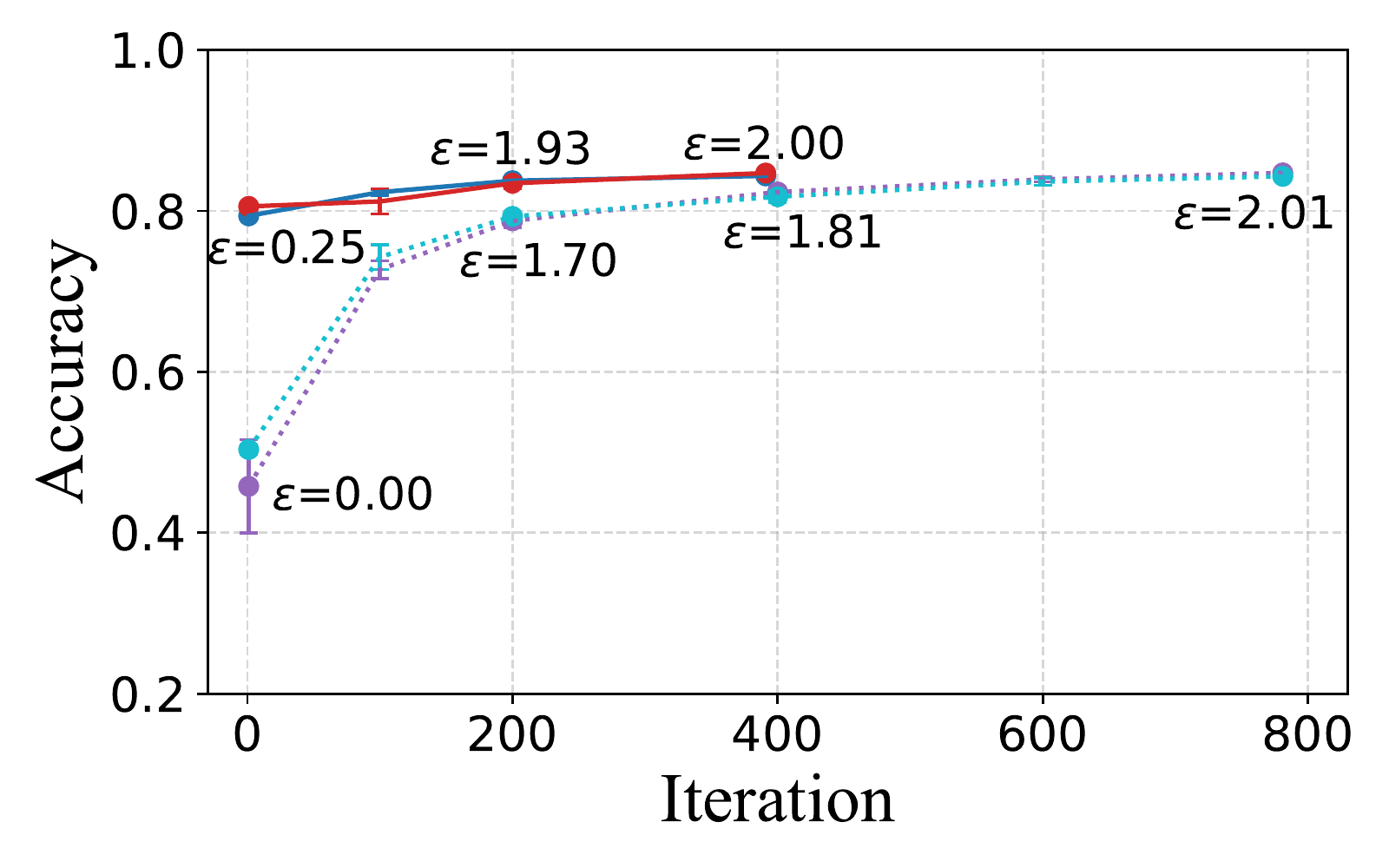}
    }
    \includegraphics[width=0.8\linewidth]{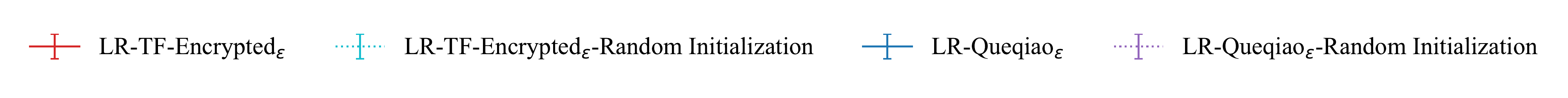}
    \caption{Convergence speed comparison of local data initialized global models and randomly initialized global models.
    }
    \label{fig:initialization}
\end{figure*}

\subsection{Effectiveness of Optimizations}\label{subsec:effectiveness_optimizations}
\noindent\textbf{Effectiveness of Feature Extraction.}
We keep other experimental settings unchanged and compare the accuracy of differentially private models trained on extracted features and raw data, respectively. For MNIST and CIFAR-10, we directly use their pixel matrices as training data. For IMDb, we perform tokenization on its textual data with BertTokenizerFast tool\footnote{\url{https://github.com/huggingface/transformers/blob/master/src/transformers/models/bert/tokenization_bert_fast.py}} to produce the training data that are represented as a matrix. 


We show the experimental results in Figure~\ref{fig:feature_extraction}. Compared with differentially private models trained on raw data, models trained on features extracted by our method can achieve higher accuracy. For CIFAR-10, which has the highest dimension (3,072) among the three datasets, the accuracy gap between models trained on extracted features and raw data is higher than 60\%. For IMDb, which has rich semantics, the models trained on raw data are almost totally random. The above results show that extracting features of input data with data-independent feature extractors can significantly improve the accuracy of classification models. 

\noindent\textbf{Effectiveness of Global Model Initialization with Local Data.}  To set the baseline, we use random variables sampled from Gaussian distribution to initialize the global model, which is a standard initialization method in the machine learning field~\cite{goodfellow2016deep}. We set the mean and the standard deviation of the Gaussian distribution as 0 and 0.2. As to the privacy parameter settings, we set $\epsilon$ as the sum of $\epsilon_1$ used in the local training phase and $\epsilon_2$ used in the global training phase, and $\delta$ as $\frac{1}{10n}$.

We show the experimental results in Figure~\ref{fig:initialization}. The experimental results demonstrate the effectiveness of our proposed global model initialization with local data method. When we start training from a randomly initialized global model, even though after twice of iterations, the accuracy of trained models is still smaller than the models trained from the models that are initialized with local data. Therefore, by consuming part of the privacy budget locally, we can significantly reduce the number of secure computations in the global model training phase and improve the efficiency of the model training process.

\subsection{Comparison of Aggregation Methods}~\label{subsec:compare_aggregation}
In this section, we compare the accuracy of the global model initialized by averaging local models and choosing the most accurate one. We consider two scenarios: independent and identically distributed (IID), and Non-IID. We randomly divide the whole dataset into three parts to simulate the IID scenario. For the Non-IID scenario, following the previous study~\cite{mcmahan2017communication}, we first sort the whole datasets by label values, then divide them into three parts. After partitioning the datasets, we train logistic regression models with the same parameter settings of Section~\ref{subsec:end-to-end} and aggregate them with different methods. 

\begin{table}[ht]
\centering
\caption{Accuracy (in \%) comparison of global models initialized by different aggregation methods. The highest accuracy values are marked in bold.}
\label{tab:aggregation_comparison}
\scalebox{0.85}{
\begin{tabular}{|c|c|c|c|}
\hline
\textbf{Dataset}                   & \textbf{Data distribution} & \textbf{Averaging Strategy} & \textbf{Accuracy Strategy} \\ \hline
\multirow{2}{*}{MNIST}    & IID               & $\textbf{89.53}\; (\pm 0.26)$ & $88.45\; (\pm 0.40)$ \\ \cline{2-4} 
                          & Non-IID           & $\textbf{50.56}\; (\pm 1.67)$ & $39.59\; (\pm 0.41)$ \\ \hline
\multirow{2}{*}{CIFAR-10} & IID               & $\textbf{83.75}\; (\pm 0.36)$ & $82.16\; (\pm 0.42)$ \\ \cline{2-4} 
                          & Non-IID           & $\textbf{65.39}\; (\pm 1.87)$ & $38.80\; (\pm 0.06)$ \\ \hline
\multirow{2}{*}{IMDb}     & IID               & $\textbf{81.11}\; (\pm 0.57)$ & $80.53\; (\pm 1.01)$ \\ \cline{2-4} 
                          & Non-IID           & $67.48\; (\pm 3.04)$ & $\textbf{79.10}\; (\pm 1.36)$ \\ \hline
\end{tabular}}
\end{table}

We show the accuracy comparison results in Table~\ref{tab:aggregation_comparison}.   
In the IID scenario, the performance of averaging local models is slightly higher than choosing the most accurate local model. In contrast, in the Non-IID scenario, these two aggregation methods have different performances on different datasets. For MNIST and CIFAR-10, the global models initialized by averaging local models are much more accurate than those initialized by choosing the most accurate local model. While for IMDb, the global models initialized by choosing the most accurate local model are much more accurate than those initialized by averaging parameters of local models. As we choose the optimal aggregation method by comparing the accuracy of candidate models in Protocol~\ref{alg:global_model_initialization}, we can initialize an accurate global model for different scenarios.

\vspace{-2mm}
\subsection{Comparison with Federated Learning and Secure Aggregation with DP}~\label{subsec:FL_secure_DP}
In this section, we compare \wqruan{PEA} with two state-of-the-art methods on federated learning and secure aggregation with DP, namely CAPC Learning (CAPC for short)~\cite{choquette-choo2021capc} and Distributed Discrete Gaussian mechanism for federated learning with secure aggregation (DDGauss for short)~\cite{kairouz2021distributed}. We directly run their open-source codes\footnote{\url{https://github.com/cleverhans-lab/capc-iclr}}$^{,}$\footnote{\url{https://github.com/google-research/federated}} \wqruan{on extracted features of MNIST and CIFAR-10 datasets} as baselines. We set privacy parameters of \tfencrypted, \queqiao, and CAPC as $\epsilon = 2, \delta = \frac{1}{10n}$. For DDGauss, we set $\epsilon$ as \wqruan{5} and $\delta$ as $\frac{1}{10n}$ as the training process of DDGauss cannot converge when we set $\epsilon$ as 2.  For other settings (e.g. types of trained models), we set them following the description of corresponding studies~\cite{choquette-choo2021capc, kairouz2021distributed}. Note that we only run experiments on MNIST and CIFAR-10 datasets because the sentiment analysis task is not supported by the open-source codes of CAPC and DDGauss.

\begin{table}[ht]
\centering
\caption{Accuracy (in \%) comparison of models trained by \tfencrypted and \queqiao and models trained by CAPC and DDGauss. We report the average results of five runs and show the standard deviations in brackets.}
\label{tab:acc_FL_MPL}
\scalebox{0.85}{
\begin{tabular}{|c|c|c|c|c|}
\hline
\multicolumn{1}{|l|}{} & \textbf{\tfencrypted} & \textbf{\queqiao}  &\textbf{ CAPC~\cite{choquette-choo2021capc} }               & \textbf{DDGauss~\cite{kairouz2021distributed}}            \\ \hline
MNIST                  & 95.14 ($\pm 0.09$)      & 94.90 ($\pm 0.14$) & \wqruan{92.52 ($\pm 0.20$)} & \wqruan{92.45 ($\pm 0.38$)} \\ \hline
CIFAR-10               & 88.99 ($\pm 0.08$)      & 88.18 ($\pm 0.33$) & \wqruan{86.70 ($\pm 0.17$)}  & \wqruan{53.31 ($\pm 2.52$)} \\ \hline
\end{tabular}
}
\end{table}

\noindent\textbf{Result.}  As is shown in Table~\ref{tab:acc_FL_MPL}, the models trained by \tfencrypted and \queqiao are all more accurate than the models trained by CAPC and DDGauss. \wqruan{ CAPC applies SMPC to use other parties' models to securely label one party's local data, thus the party can retrain its model on the augmented dataset. DDGauss applies SMPC to securely aggregate noisy gradients from parties. These two methods inevitably miss some information about the original dataset. In contrast, we apply SMPC protocols to simulate a trusted server that could virtually hold all data from parties to train target models~\cite{lindell2017simulate}. Therefore, even though CAPC and DDGauss can efficiently support larger numbers of parties (50 and 500 in our experiments) than PEA, the models trained by CAPC and DDGauss suffer more accuracy loss than the models trained by \tfencrypted and \queqiao.} \newchange{Meanwhile, it is worth noting that CAPC applies DP to protect parties' private information during the data labeling process and does not consider the membership inference attacks from end-users. We further discuss this point in Section~\ref{related_work}.} 

\subsection{Accuracy Evaluation of Models Trained by MPL}\label{subsec:effectiveness_MPL}

In this section, we conduct two experiments to evaluate the accuracy of models trained by MPL: (1) comparing the accuracy of global models with the accuracy of local models, and (2) evaluating the impact of SMPC on the accuracy of trained models. For the first experiment, we randomly divide the whole dataset into 3, 5, and 10 disjoint parts to simulate scenarios with different numbers of parties. After extracting features, we train differentially private logistic regression models on one plaintext subset and the logically global dataset with the same privacy parameters respectively. For the second experiment, we fix other experiment settings and compare the accuracy of models trained on the global dataset in the plaintext form and the logically global dataset merged by SMPC.

\noindent\textbf{Results.} We show the experimental results in Table~\ref{tab:effectiveness_MPL_CIFAR} and Table~\ref{tab:ablation} respectively. For the first experiment, due to the space limitation, we only show the results of CIFAR-10 and show the results of other datasets in Appendix~\ref{missing_res_MPL}. As is shown in Table~\ref{tab:effectiveness_MPL_CIFAR}, the models globally trained by \tfencrypted and \queqiao are all more accurate than the models locally trained on one subset of the whole dataset. Moreover, as the number of parties increases, the accuracy gap between locally trained and globally trained models becomes larger. There are mainly two reasons behind the above result. First, more training data provide more features, thus producing a more accurate model. Second, a larger training dataset implies a smaller sample probability, thus requiring less random noise to achieve the same privacy protection level. 

For the second experiment, as is shown in Table~\ref{tab:ablation}, SMPC protocols have a negligible impact (i.e. less than 0.9\%) on the accuracy of trained models. This is because we apply SMPC protocols to simulate a trusted server that could virtually hold all data to train the target model~\cite{lindell2017simulate}. The possible impact of SMPC mainly comes from two factors: (1) we perform the computation over fixed-point numbers. However, they can only express the decimal number with limited precision, which brings a negligible accuracy loss. (2) we use linear piecewise functions to approximate non-linear activation functions (e.g. sigmoid function), which would bring another negligible accuracy loss. In summary, MPL can significantly improve the accuracy of differentially private models.

\begin{table} 
\footnotesize
\centering
\caption{Accuracy (in \%) comparison of models  securely trained on the logically global dataset and models locally trained on one subset of the whole dataset for CIFAR-10. We report the average results of five runs and show the standard deviations in brackets. Note that we only show the 3-party result of \tfencrypted because its back-end SMPC protocol (i.e. \texttt{ABY3}) is tailored for the 3-party scenario. }
\label{tab:effectiveness_MPL_CIFAR}
\scalebox{0.85}{
\begin{tabular}{|c|c|c|c|c|}
\hline
\textbf{\#Party}    & \textbf{$\epsilon$}  & \textbf{Local party}  & \textbf{\tfencrypted}  & \textbf{\queqiao}     \\ \hline
\multirow{3}{*}{3}  & 0.25                 & $82.65\; (\pm 0.30)$  & $84.03\; (\pm 0.28)$   & $84.35\; (\pm 0.51)$  \\ \cline{2-5} 
                    & 1                    & $86.43\; (\pm 0.21)$  & $88.57\; (\pm 0.11)$   & $87.79\; (\pm 0.27)$  \\ \cline{2-5} 
                    & 2                    & $87.15\; (\pm 0.13)$  & $88.99\; (\pm 0.08)$   & $88.18\; (\pm 0.33)$  \\ \hline
\multirow{3}{*}{5}  & 0.25                 & $79.84\; (\pm 0.55)$  & -                      & $82.09\; (\pm 0.38)$  \\ \cline{2-5} 
                    & 1                    & $85.70\; (\pm 0.37)$  & -                      & $88.05\; (\pm 0.25)$  \\ \cline{2-5} 
                    & 2                    & $86.07\; (\pm 0.09)$  & -                      & $88.34\; (\pm 0.24)$  \\ \hline
\multirow{3}{*}{10} & 0.25                 & $69.49\; (\pm 1.14)$  & -                      & $77.71\; (\pm 0.39)$  \\ \cline{2-5} 
                    & 1                    & $81.32\; (\pm 1.01)$  & -                      & $88.05\; (\pm 0.39)$  \\ \cline{2-5} 
                    & 2                    & $83.35\; (\pm 0.52)$  & -                      & $88.09\; (\pm 0.14)$  \\ \hline
\end{tabular}}
\end{table}

    \begin{table}[ht]
        \centering
   \caption{  Accuracy (in \%) comparison of differentially private models trained by MPL on distributed data and models centrally trained on the global dataset in plaintext. We report the average results of five runs and show the standard deviations in brackets. We set $\epsilon$ as 2 when training the models.}
   \scalebox{0.85}{
    \begin{tabular}{|c|c|c|c|}
\hline
\multicolumn{1}{|l|}{} & \textbf{\tfencrypted} &\textbf{\queqiao}  & \textbf{Plaintext}   \\ \hline
MNIST                   & 95.14 ($\pm 0.09$)    & 94.90 ($\pm 0.14$)     & 
95.35  ($\pm 0.07$)    \\ \hline
CIFAR-10               & 88.99 ($\pm 0.08$)  & 88.18 ($\pm 0.33$)       & 89.00 ($\pm 0.13$)     \\ \hline
IMDb                  & 84.70   ($\pm 0.44$)  & 84.35 ($\pm 0.17$)     &85.19 ($\pm 0.14$)     \\ \hline
\end{tabular}}
\label{tab:ablation}
    \end{table}


\section{Related Work}~\label{related_work}

\noindent\textbf{Multi-party Learning.} In recent years, implementing privacy-preserving machine learning based on SMPC protocols has become a hot topic in the security field. Nikolaenko et al.~\cite{6547119} combined homomorphic encryption and garbled circuits to train ridge regression models securely. Mohassel and Zhang~\cite{DBLP:conf/sp/MohasselZ17} first applied additive secret sharing to train neural network models securely. After that, many following studies were proposed to improve the efficiency of the training process~\cite{mohassel2018aby3,cryptGPU, wagh2020falcon}, enhance security models~\cite{ fantastic-four}, or support the training of more complex models~\cite{wagh2020falcon, cryptGPU}. 
For example, \texttt{CryptGPU}~\cite{cryptGPU} and \texttt{Falcon}~\cite{wagh2020falcon} improved the efficiency of computation primitives and added more computation primitives to support the training of deep neural network models (e.g. VGG-16). 
Different from previous studies, which optimize the efficiency of computation primitives involved in the model training process, we try to optimize the training of MPL from the model training process aspect, i.e. reducing the secure computations required for the training process.

\noindent\textbf{Differentially Private Machine Learning.} There are mainly three paradigms in differentially private machine learning algorithms: objective perturbation, output perturbation, and gradient perturbation. Objective perturbation adds random noise to loss functions as a regularization term. Output perturbation~\cite{chaudhuri2011differentially, wu2017bolt} adds random noise to trained models. They both assume that loss functions are convex and continuous to preserve the rigorous privacy guarantees, which significantly limits the application scenarios of these two paradigms. Therefore, we choose to enforce DPSGD~\cite{deepdp,bassily2014private,song2013stochastic}, which follows the gradient perturbation paradigm and has no assumption on the loss functions, in secret sharing-based MPL frameworks to enhance the privacy of these frameworks.

Recently, there have been several studies~\cite{papernot2020making,dimension_reduction,tramer2021differentially} that utilize priors knowledge to improve the utility of DPSGD. Papernot et al.~\cite{papernot2020making} improved the accuracy of models trained by DPSGD by using public data to initialize model parameters. Yu et al.~\cite{dimension_reduction}  utilized few public data to project the high-dimension gradient vectors to a low-dimension space, thus reducing the noise added in the training process. Tramèr and Boneh~\cite{tramer2021differentially} showed that applying heuristic rules or knowledge transferred from public data to extract features of input data could significantly reduce the accuracy loss brought by DPSGD on image classification tasks. \newchange{However, their feature extraction method might be less effective in the multi-party setting. We empirically show that our feature extraction method is more suitable for the multi-party setting than that proposed by Tramèr and Boneh~\cite{tramer2021differentially} in Appendix~\ref{acc_Feature}. Meanwhile, we consider the sentiment analysis, which is an important task of NLP.}


\noindent\textbf{Intersection of SMPC and DP.}
Because the security models of SMPC and DP are complementary, i.e. SMPC guarantees the security of the computation process and DP preserves output privacy, there have been many studies on the intersection of SMPC and DP. Wagh et al.~\cite{DP_Crypt} provided an overview of recent studies on the intersection of SMPC and DP. Pettai et al.~\cite{acsac15} applied secret sharing protocols to improve the utility of differentially privacy aggregation queries on sensitive data. Next, He et al.~\cite{10.1145/3133956.3134030} combined secret sharing protocols and DP to securely link similar records in different databases. In recent years, as many strict privacy protection regulations have been published, some related studies~\cite{dp_smpc_heavy_hitter,dp_smpc_median,cryptepsilon} have been proposed to securely collect and analyze sensitive data in a differentially private manner. While there are several studies on secure and differentially private statistical analyses, such as aggregation queries~\cite{acsac15, cryptepsilon}, median~\cite{dp_smpc_median} or heavy hitters~\cite{dp_smpc_heavy_hitter}, there still lacks a combination of differentially private machine learning algorithms and MPL.

\noindent\textbf{Federated Learning and Secure Aggregation with DP.} In addition to our proposed secure DPSGD protocol, there is another routine of studies on securely training differentially private machine learning models on distributed data, namely federated learning and secure aggregation with DP~\cite{chase2017private,choquette-choo2021capc, kairouz2021distributed}. Chase et al.~\cite{chase2017private} applied garbled circuits~\cite{ref_yao} to securely aggregate noisy gradient vectors from parties. Choquette-Choo et al. proposed CAPC~\cite{choquette-choo2021capc} to extend PATE~\cite{pate} to the distributed setting by combining homomorphic encryption~\cite{yi2014homomorphic} and garbled circuits~\cite{ref_yao}. 
Kairouz et al.~\cite{kairouz2021distributed} extended the discrete gaussian mechanism~\cite{canonne2020discrete} to the distributed setting. \wqruan{These studies apply SMPC protocols to securely aggregate the intermediate information (i.e. noisy labels or noisy gradient vectors) of the training process, which implies relatively higher efficiency and scalability.}  However, transferring intermediate information inevitably causes considerable accuracy loss to trained models~\cite{yang2019federated}, especially when the data are not independent and identically distributed among parties~\cite{yang2019federated}. As a result, when the number of parties is small, our proposed secure DPSGD protocol is more suitable than these methods.

\newchange{In addition, the security model of PEA concerns membership inference attacks from end-users, which are not concerned by the security model of CAPC~\cite{choquette-choo2021capc}. The training process in CAPC consists of three steps: (1) each party trains its local model; (2) a querying party employs CAPC protocol to confidentially and privately query other parties to obtain (query data, label) pairs; (3) the querying party retrains an improved local model with the pairs.  The CAPC protocol guarantees both the confidentiality of the query data and the privacy of the local models in Step (2). However, they assume that the retrained model is only used by the querying party and do not consider the membership inference attacks from other users.} 


\section{Discussion and Future Work}\label{discussion}
\noindent\textbf{Side-channel Attacks against Differentially Private Mechanisms.}
There have been several side-channel attacks~\cite{least_bit,exp_dp_2} on the implementations of DP mechanisms.  Mironov~\cite{least_bit} proposed an attack that exploits the limited precision of floating-point number arithmetic to destroy the privacy guarantee of Laplacian mechanisms. While these side-channel attacks severely break the privacy guarantee of DP mechanisms in practice, they mainly target the statistical queries on databases, where random noises are added once to query results. In DPSGD, the random noises are iteratively added in the training process, and only the trained model is released. Without the knowledge of the initial model and the information of intermediate gradient vectors, attackers cannot exploit side-channel attacks to destroy the privacy guarantee of our proposed secure DPSGD protocol.

\noindent\textbf{Feature Extraction for Other Types of Data.} 
Following previous studies~\cite{DBLP:conf/sp/MohasselZ17, cryptGPU,Bu2020Deep}, we consider the image data and textual data in our experiments. However, our proposed optimization methods are also applicable to other types of data. There have emerged several foundation models that can effectively extract the features of other types of data (e.g.  protein sequences~\cite{jumper2021highly}) to finish downstream tasks. Therefore, our proposed optimization methods are general enough to support various types of input data. 

\noindent\textbf{Recent Backdoor Attacks on Foundation Models.}~\label{subsec:backdoor_attacks}
Recently, Jia et al.~\cite{jia2022badencoder} and Zhang et al.~\cite{9581257} proposed two backdoor attacks on foundation models. Here, attackers inject backdoors into downstream classification models by adding adversarial samples in the training process of foundation models. However, these two attacks assume that attackers hold a few training samples for the target tasks. With the strict protection of highly sensitive training data  (e.g. medical images) in the context of MPL, attackers would be difficult to obtain these data samples, thus leading to little privacy risk to the trained models. Besides, to further defend the backdoor attacks on foundation models, the owners of trained models can employ recently proposed defenses (e.g. MNTD~\cite{xu2021detecting}) to detect the backdoors hidden in the trained models.


\noindent\textbf{Future Work.}~\label{subsec:future_work}
We will integrate recent optimizations~\cite{ dimension_reduction} on DPSGD into existing MPL frameworks to further reduce the accuracy loss brought by DP. Concretely, we first project high-dimension gradient vectors to a low-dimension space with a few auxiliary public data. Then we perturb the low-dimension gradient vectors and project perturbed gradient vectors to the original high-dimension space. In this way, we can reduce the random noise added in the model training process, then improve the accuracy of the trained differentially private model.  Moreover, we will implement the state-of-the-art defenses (e.g. prediction poisoning~\cite{Orekondy2020Prediction} and prediction purification~\cite{yang2020defending}) against model stealing attacks~\cite{model_extraction_2016} and model inversion attacks~\cite{model_inversion_2015}, which cannot be defended by DPSGD, in MPL frameworks to further enhance their privacy protection. 



\section{Conclusion}\label{conclusion}
In this paper, we propose \wqruan{PEA}, which can help secret sharing-based MPL frameworks to securely and efficiently train a differentially private machine learning model with little accuracy loss. After implementing \wqruan{PEA} in two open-source MPL frameworks: \texttt{TF-Encrypted} and \texttt{Queqiao}, we conduct experiments on three datasets: MNIST, CIFAR-10, and IMDb. The experimental results demonstrate the efficiency and effectiveness of \wqruan{PEA}. For all three datasets, \tfencrypted and \queqiao can train accurate differentially private classification models within much fewer iterations than differentially private deep neural network models. In particular, compared with \texttt{CryptGPU}, \tfencrypted only requires less than 1\% (7 minutes vs. 16 hours) of its time to train a classification model for CIFAR-10 with the same accuracy. In conclusion, with \wqruan{PEA}, multiple parties can balance the 3-way trade-off between privacy, efficiency, and accuracy in the model training process of MPL.


\section*{Acknowledgment}
This paper is supported by NSFC (No. U1836207, 62172100) and STCSM (No. 21511101600). We thank Ninghui Li, Haoqi Wu, Guopeng Lin, Xinyu Tu, Shuyu Chen, and all reviewers for their insightful comments. Weili Han is the corresponding author.


%
\normalem
\bibliographystyle{plain}

\smallskip
\appendices
\vspace{-3mm}
\section{Missing Proofs}\label{Missing Proof}
\noindent \textbf{ Lemma~\ref{parallel_composition}.} For $n$ mechanisms $M_1, M_2, \cdots, M_n$  whose inputs are disjoint datasets $D_1, D_2, \cdots, D_n$, if $M_i$ satisfies ($\epsilon_i, \delta_i$)-differential privacy, the combination of them satisfies ($\epsilon$, $\delta$)-differential privacy  for the dataset  $D = D_1 \cup D_2 \cup \cdots \cup D_n$ with $\epsilon = Max(\epsilon_1, \epsilon_2, \cdots, \epsilon_n)$, $\delta = Max(\delta_1, \delta_2, \cdots, \delta_n)$. 

\begin{proof}
We first redefine ($\epsilon, \delta$)-differential privacy in a new form based on Definition~\ref{new_def_dp}~\cite{parallel_composition}.

\begin{myDef}\label{new_def_dp}
For two datasets $D$ and $D'$, we say a random mechanism M that outputs \begin{math}r \in R^p\end{math} satisfies ($\epsilon, \delta$)-differential privacy if, for any subset $S \subseteq  R^p$,
$Pr(M(D) \in S) \leq e^{\epsilon * |D\oplus D'|} \cdot Pr (M(D') \in S) +  \delta * |D\oplus D'|$
\end{myDef}
\noindent where $D\oplus D'$ is the symmetric difference between $D$ and $D'$. Then for $D_1, \cdots, D_n$ and $D'_1, \cdots, D'_n$, let $M$ be the combination of $M_1 \cdots, M_n$, the probability that $M$ outputs $r$ is
    $Pr(M(D) = r) = \prod_{i = 1}^n Pr(M_i(D_i) = r_i) $. Meanwhile,
as $D$ and $D'$ are neighboring datasets, there is one $|D_i\oplus D'_{i}|$ larger than 0 and is equal to $|D\oplus D'|$. Without loss of generality, we assume $|D_k\oplus D'_{k}| > 0$. Therefore, 
\begin{small}
\begin{align*}
&\quad \prod_{i = 1}^n Pr(M_i(D_i) = r_i)\\
&\leq \prod_{i = 1}^n (e^{\epsilon_i * |D_i\oplus D'_i|} Pr(M_i(D'_i) = r_i) + \delta_i * |D_i\oplus D'_i|)\\
& = (\prod_{i = 1, i \neq k}^n Pr(M_i(D'_i) = r_i)) *(e^{\epsilon_k * |D_k\oplus D'_k|}Pr(M_k(D'_k) = r_k)\\
&\quad + \delta_k *|D_k\oplus D'_k|)\\
&\leq e^{Max(\epsilon_1, \cdots, \epsilon_n) * |D_k\oplus D'_k|}\prod_{i = 1}^n Pr(M_i(D'_i) = r_i)\\
&\quad + (\prod_{i = 1, i \neq k}^n Pr(M_i(D'_i) = r_i)) * Max(\delta_1, \cdots, \delta_n) * |D_k\oplus D'_k|\\
&\leq  e^{Max(\epsilon_1, \cdots, \epsilon_n) * |D\oplus D'|}\prod_{i = 1}^n Pr(M_i(D'_i) = r_i)\\
&\quad +Max(\delta_1, \cdots, \delta_n)* |D\oplus D'|
\end{align*}
\end{small}
It completes the proof.
\end{proof}

\noindent \textbf{ Lemma~\ref{lemma:error_bound}.} For any $x'$ $ \in [0.5, 1),  (0.8277x'^2-2.046x'+2.223 - 0.0048) -\frac{1}{\sqrt{x'}} < 0$.

\begin{proof}
Let $f(x') = (0.8277x'^2-2.046x'+2.223-0.0048) -\frac{1}{\sqrt{x'}}$. As function $f$ is differentiable on interval $[0.5, 1)$, we can get its first-order derivative function as  $f'(x') = 1.6554x'-2.046 + \frac{1}{2\sqrt{x'^{3}}} $ and second-order derivative function as $f''(x') = 1.6554 - \frac{3}{4\sqrt{x'^{5}}}$. Because $f''(x')$ is monotonically increasing on $[0.5, 1)$ and has one zero point $x'_0 = 0.7286$, $f'(x')$ is monotonically decreasing on $[0.5, 0.7286)$ and monotonically increasing on $[0.7286, 1)$. Then as $f'(0.7286) = -0.0359 < 0$ and $f'(x')$ is larger than 0 on two endpoints $0.5$ and $1$, $f'(x')$ has two zero points on $[0.5, 1)$: $x'_1 = 0.6257$ and $x'_2 = 0.8517$. Therefore, $f(x')$ is monotonically increasing on $[0.5, 0.6257)$, monotonically decreasing on $[0.6257, 0.8517)$, and monotonically increasing on $[0.8517, 1)$. Finally, for any $x' \in [0.5, 1)$, $f(x') < \max\{f(0.6257), f(1)\} = -0.0001 < 0$. 
\end{proof}

\vspace{-1mm}
\noindent \textbf{ Theorem~\ref{theo:secure_dp_sgd}.}   For any $\epsilon \leq 2\log(1/\delta)$ and $\delta \in (0, 1)$,  Protocol~\ref{alg:secure_dp_sgd} satisfies ($\epsilon, \delta$)-differential privacy.

\begin{proof}
We first introduce a differential privacy concept, namely R\'enyi differential privacy (RDP)~\cite{renyi}, to tightly analyze the privacy budget of DPSGD in a simple way.

\begin{myDef}
\textbf{($\lambda,\gamma$)-R\'enyi Differential Privacy~\cite{renyi}.}  For a random mechanism M whose input is \begin{math}D\end{math} and output \begin{math}r \in R^p\end{math}, we say M satisfies  \begin{math}(\lambda,\gamma)\end{math}-RDP if, for any two neighboring datasets $D, D'$, it holds that,
\begin{center}
\scalebox{0.9}{
    \begin{math}
    D_{\lambda}(M(D)\|M(D')) \leq \gamma
    \end{math}
}
\end{center}
\end{myDef}
\noindent where $D_{\lambda}(M(D)\|M(D'))$ is $\lambda$-R\'enyi divergence between the distributions of $M(D)$ and $M(D')$. Then we show the proof of Theorem~\ref{theo:secure_dp_sgd}.

According to Lemma~\ref{lemma:error_bound}, the maximum difference between the approximated $\frac{1}{\sqrt{x'}}$ and the true $\frac{1}{\sqrt{x'}}$ is smaller than 0. Thus Protocol~\ref{alg:secure_inversesqrt} can ensure that for each iteration, the $L_2$ sensitivity of Protocol~\ref{alg:secure_dp_sgd} is smaller than $C$. Then each iteration of Protocol~\ref{alg:secure_dp_sgd} satisfies ($\lambda, \frac{\lambda C^{2}}{2\sigma^{2}}$)-RDP according to Corollary 3 of \cite{renyi}. Next, according to Proposition 1 of \cite{renyi}, Protocol~\ref{alg:secure_dp_sgd} satisfies ($\lambda,  T\frac{\lambda C^{2}}{2\sigma^{2}}$)-RDP. Finally, according to Proposition 3 of \cite{renyi}, in order to ensure ($\epsilon, \delta$)-differential privacy, we should ensure that
\begin{center}
    \begin{math}
    \frac{TC^{2}\lambda}{2\sigma^{2}} + \frac{\log(\frac{1}{\delta})}{\lambda - 1} \leq \epsilon
    \end{math}
\end{center}
\noindent We set $\lambda = 1 + \frac{2\log(\frac{1}{\delta})}{\epsilon}$, thus we should ensure that
\begin{center}
    \begin{math}
    \sigma^{2} \geq \frac{TC^{2}(\epsilon+2\log(\frac{1}{\delta}))}{\epsilon^{2}} 
    \end{math}
\end{center}
\noindent Meanwhile, as we set in Protocol~\ref{alg:secure_dp_sgd}
\begin{center}
    \begin{math}
    \sigma^2 \geq \frac{TC^{2}(4\log(\frac{1}{\delta}))}{\epsilon^{2}} \geq \frac{TC^{2}(\epsilon+2\log(\frac{1}{\delta}))}{\epsilon^{2}} 
    \end{math}
\end{center}
Therefore, Protocol~\ref{alg:secure_dp_sgd} satisfies ($\epsilon, \delta$)-differential privacy.
\end{proof}

\begin{table}
\centering
\caption{Hyperparameter settings of local model training.}
\label{tab:local_model_paramters}
\scalebox{0.85}{
\begin{tabular}{|c|c|c|c|c|c|c|}
\hline
\textbf{Datasets} & \textbf{$\epsilon$} & \textbf{$\delta$} & \textbf{Clip bound} & \textbf{\#Epoch} & \textbf{Batch size} & \textbf{LR} \\ \hline
MNIST    & 0.25                        &      $2\times10^{-7}$                 &     1           &    3       &        128    &          0.1     \\ \hline
CIFAR-10 &    0.25                     &    $1.7\times10^{-7}$                     &   1          &    1       &    128        &     0.1          \\ \hline
IMDb     & 0.25                        &      $4\times10^{-7}$                   &       1         &    3       &    128       &     0.1          \\ \hline
\end{tabular}}
\end{table}

\begin{table}
\centering

\caption{Potential hyperparameters involved in the grid search procedure of baselines training.}
\label{tab:baseline_grid_search_paramters}
\scalebox{0.85}{
\begin{tabular}{|c|c|}
\hline
\textbf{Hyperparameter}          &   \textbf{Potential values}                            \\  \hline
Clipping bound          &   1, 2, 5                                     \\  \hline
Batch size              &   64, 128, 256, 512                               \\  \hline
Learning rate schedule  &   Constant (lr=0.1), OneCycleLR (max\_lr=0.1) \\  \hline

\end{tabular}}
\end{table}

\noindent\textbf{Theorem~\ref{theo:model_initialization}.} Protocol~\ref{alg:global_model_initialization} satisfies ($\epsilon_1, \delta_1$)-differential privacy.
\begin{proof}
 Each local dataset $D_i$ held by $P_i$ can be viewed as a subset of the logically global dataset. Meanwhile, as different parties hold different data, their local datasets are disjoint. In addition, the aggregation methods are data-independent. According to the post-process immunity property of differential privacy, the aggregation has no impact on the privacy guarantee. Therefore, we only need to consider the privacy guarantee of local model training. 
 As all parties privately train their models with privacy parameters ($\epsilon_1, \delta_1$), we can obtain the privacy guarantee of Protocol~\ref{alg:global_model_initialization} by directly applying Lemma~\ref{parallel_composition}. 
\end{proof}

\vspace{-2mm}
\section{Experiment Settings}\label{parameter setting}

\begin{table}
\centering

\caption{Hyperparameter settings of baseline model training.}
\label{tab:baseline_paramters}
\scalebox{0.9}{
\begin{tabular}{|c|c|c|c|}
\hline
\textbf{Datasets} & \textbf{Clipping bound} & \textbf{Batch size} & \textbf{LR} \\ \hline
MNIST    &     1          &    128     & Constant (lr=0.1)     \\ \hline
CIFAR-10 &     1          &    512     & OneCycleLR (max\_lr=0.1)  \\ \hline
IMDb     &     2          &    64      & OneCycleLR (max\_lr=0.1)   \\ \hline
\end{tabular}}
\end{table}

In this section, we show the hyperparameter settings of experiments. The hyperparamters of local model training are shown in Table~\ref{tab:local_model_paramters}.  The potential hyperparameters involved in the grid search procedure are shown in Table~\ref{tab:baseline_grid_search_paramters}.  We also show the parameters found by the grid search procedure in Table~\ref{tab:baseline_paramters}. Note that we run the baseline training process of MNIST, CIFAR-10, and IMDb for 20, 80, 20 epochs, respectively.

\section{Security and Communication Complexity Analysis of Protocols}\label{sec:secrity_complexity_analysis}

\subsection{Analysis of Protocol~\ref{alg:secure_inversesqrt}}~\label{sec:analysis_inversesquareroot}
\vspace{-2mm}

\noindent\textbf{Security.} As Protocol~\ref{alg:secure_inversesqrt} only contains the cryptographic primitives whose security has been proven in previous studies~\cite{div2mp,10.1007/978-3-642-15317-4_13} and the rest of computation is performed locally. Therefore, as long as the security of these cryptographic primitives holds, Protocol~\ref{alg:secure_inversesqrt} is secure.

\noindent\textbf{Communication Complexity.} In Protocol~\ref{alg:secure_inversesqrt}, the computation of $2^{f-\frac{exp}{2}}$ and the evaluation of approximated polynomial can be parallelized. Therefore, in the critical path of Protocol~\ref{alg:secure_inversesqrt}, there are four $\mathsf{Multiplication}$ and six $\mathsf{Trunc}$ invocations, two $\mathsf{Bit\_Dec}$ invocations, one $\mathsf{Mod2}$, $\mathsf{SufOr}$ and $\mathsf{PreMulC}$ invocation, respectively. According to Table~2 and Table~3 of \cite{div2mp}, Table~4 of \cite{10.1007/978-3-642-15317-4_13}, the total communication round number and the communication complexity of Protocol~\ref{alg:secure_inversesqrt} is 15, $O(mk^2)$ bits respectively. Meanwhile, the offline  communication round number and the communication complexity of Protocol~\ref{alg:secure_inversesqrt} is 1, $ O(mk^2+mfk)$ bits respectively. Note that the communication round is 1 because the offline computations have no mutual dependence and can all be executed in parallel.

\subsection{Analysis of Protocol~\ref{alg:secure_dp_sgd}}~\label{sec:analysis_secure_dpsgd}
\vspace{-2mm}

\noindent\textbf{Security.} In addition to the invocations of Protocol~\ref{alg:secure_inversesqrt} and Protocol~\ref{alg:secure_noise_generation}, the other secure computation primitives used in Protocol~\ref{alg:secure_dp_sgd} (i.e. comparison, multiplication and subtraction) all have the standard protocols~\cite{ref_bgw,ref_gmw,div2mp}, which have been proven secure by previous studies~\cite{ref_bgw, ref_gmw, div2mp}. Meanwhile, because the security of Protocol~\ref{alg:secure_inversesqrt} and Protocol~\ref{alg:secure_noise_generation} has been stated in Section~\ref{Sec:Inverse} and Section~\ref{Sec:Noise_Gen}, as long as our security model holds, Protocol~\ref{alg:secure_dp_sgd} is secure.

\noindent\textbf{Communication Complexity.} In the online phase, for each batch with batch size $B$, in addition to invoke two $\mathsf{Trunc}$, four $\mathsf{Multiplication}$, one Protocol~\ref{alg:secure_inversesqrt} and one $\mathsf{Comparison}$ $B$ times in parallel, Protocol~\ref{alg:secure_dp_sgd} contains one extra $\mathsf{Trunc}$ to update the model parameters. Therefore, according to Table 2 of \cite{div2mp}, when we execute $\mathsf{Multiplication}$ and $\mathsf{Trunc}$ in parallel, the communication round number and communication complexity of Protocol~\ref{alg:secure_dp_sgd} is 22 and $ O(mBk^2+mpBk)$ bits. Note that we consider a linear model here. For complex models with multiple layers, the communication round number keeps unchanged because gradient clipping and perturbation can be performed in parallel, while the communication complexity linearly increases as the number of parameters increases. As to the offline communication complexity, according to Table 2 of \cite{div2mp}, the offline communication round number and offline communication complexity of Protocol~\ref{alg:secure_dp_sgd} is 1 and $ O(mBk^2+mpfBk)$ bits.

\subsection{Analysis of Protocol~\ref{alg:global_model_initialization}}~\label{sec:analysis_initialization}
\vspace{-2mm}

\noindent\textbf{Security.} When parties train their local models, they only perform computations on their local data. As to the aggregation phase, the model aggregation and accuracy evaluation are completed with cryptographic primitives introduced in Section~\ref{pre:smpc}. The only information revealed to parties is the accuracy of candidate models. As these candidate models are protected by DP and testing data are not visible to parties, the accuracy does not leak private information. Therefore, as long as the security of these cryptographic primitives holds, Protocol~\ref{alg:global_model_initialization} is secure. 


\noindent\textbf{Communication Complexity.} In order to analyze the communication complexity of Protocol~\ref{alg:global_model_initialization}, we first introduce two aggregation methods of Section~\ref{subsec:initialization}. We show the details of the averaging strategy and the accuracy strategy in Protocol~\ref{alg:average_aggregation} and Protocol~\ref{alg:maximum_aggregation} respectively. We then analyze the communication complexity of each aggregation method.
\begin{algorithm}
\small
\caption{Averaging Strategy }
\label{alg:average_aggregation}
\begin{algorithmic}[1]
\REQUIRE   $P_i$ holds shares of local models $\share{\theta^{l}_{1}}_i, \share{\theta^{l}_{2}}_i,\cdots, \share{\theta^{l}_{m}}_i$.
\ENSURE  $P_i$ obtains the share of the average of local models $\share{\theta_{a}}_i$;
\STATE $P_i$ compute $\share{\theta_{a}}_i = \sum_{j = 1}^{m} \share{\theta^{l}_j}_{i} / m$;
\STATE $P_i$ obtains $\share{\theta_{a}}_i$ as the share of the average of local models;
\end{algorithmic}
\end{algorithm}
\begin{algorithm}
\small
\caption{Accuracy Strategy }
\label{alg:maximum_aggregation}

\begin{algorithmic}[1]
\REQUIRE $P_i$ holds shares of local models $\share{\theta^{l}_{1}}_i, \share{\theta^{l}_{2}}_i,\cdots, \share{\theta^{l}_{m}}_i$.
\ENSURE $P_i$ obtains the share of the most accurate local model $\share{\theta_{m}}_i$;
\FOR{Each $\theta^{l}_i \in \{\theta^{l}_1, \theta^{l}_2, \cdots, \theta^{l}_m \}$}
\STATE All parties collaboratively evaluate the accuracy of $\theta^{l}_i$ as $Accuracy_{i}$;
\IF{$Accuracy_{i} \geq$  $Accuracy_m$}
\STATE $\share{\theta_{m}} = \share{\theta^{l}_i}$;
\STATE $Accuracy_m = Accuracy_{i}$;
\ENDIF
\ENDFOR
\STATE $P_i$ obtains $\share{\theta_{m}}_i$ as the share of the most accurate local model;
\end{algorithmic}
\end{algorithm}
We first analyze the communication complexity of Protocol~\ref{alg:average_aggregation} and Protocol~\ref{alg:maximum_aggregation}. Protocol~\ref{alg:average_aggregation} only includes a constant division computation and performs one $\mathsf{Trunc}$ computation. Hence, Protocol~\ref{alg:average_aggregation} requires 2 communication rounds and $O(mpk)$ bits communication. For Protocol~\ref{alg:maximum_aggregation}, the main overhead is from the accuracy evaluation of each local model. Assuming the size of the testing dataset is $n_t$, the communication round complexity of Protocol~\ref{alg:maximum_aggregation} is $O(m)$ and the communication complexity is $O(m^2n_tpk^2)$, where the constant depends on the complexity of the trained model. 

We then analyze the communication complexity of Protocol~\ref{alg:global_model_initialization} introduced in Section~\ref{subsec:initialization}. Besides the invocations of Protocol~\ref{alg:average_aggregation} and Protocol~\ref{alg:maximum_aggregation}, Protocol~\ref{alg:global_model_initialization} evaluates the accuracy of two candidate models. Thus its communication round complexity is $O(m)$ and communication complexity is $O(m^2n_tpk^2)$. We do not analyze the offline communication complexity of the above three protocols because they do not have offline phase.

\begin{table}
\centering
\caption{Accuracy (in \%) comparison of models  securely trained on the logically global dataset and models locally trained on one subset of the whole dataset for MNIST. We report the average results of five runs and show the standard deviations in brackets.}
\label{tab:effectiveness_MPL_MNIST}
\scalebox{0.85}{
\begin{tabular}{|c|c|c|c|c|}
\hline
\textbf{\#Party}          & \textbf{$\epsilon$} & \textbf{Local party} & \textbf{\tfencrypted} & \textbf{\queqiao} \\ \hline
\multirow{3}{*}{3-party}  & 0.25                & $87.23\; (\pm 0.34)$ & $89.36\; (\pm 0.25)$                 & $89.50\; (\pm 0.22)$             \\ \cline{2-5} 
                          & 1                   & $93.56\; (\pm 0.22)$ & $95.03\; (\pm 0.11)$                 & $94.04\; (\pm 0.21)$             \\ \cline{2-5} 
                          & 2                   & $94.25\; (\pm 0.15)$ & $95.14\; (\pm 0.09)$                 & $94.90\; (\pm 0.14)$             \\ \hline
\multirow{3}{*}{5-party}  & 0.25                & $86.92\; (\pm 0.39)$ & -                                    & $88.18\; (\pm 0.21)$             \\ \cline{2-5} 
                          & 1                   & $92.33\; (\pm 0.29)$ & -                                    & $93.99\; (\pm 0.09)$             \\ \cline{2-5} 
                          & 2                   & $93.92\; (\pm 0.24)$ & -                                    & $95.13\; (\pm 0.19)$             \\ \hline
\multirow{3}{*}{10-party} & 0.25                & $79.23\; (\pm 2.23)$ & -                                    & $85.60\; (\pm 0.58)$             \\ \cline{2-5} 
                          & 1                   & $89.25\; (\pm 0.37)$ & -                                    & $94.32\; (\pm 0.31)$             \\ \cline{2-5} 
                          & 2                   & $92.96\; (\pm 0.21)$ & -                                    & $95.35\; (\pm 0.15)$             \\ \hline
\end{tabular}}
\end{table}

\begin{table}[ht]
\centering
\caption{Accuracy (in \%) comparison of models  securely trained on the logically global dataset and models locally trained on one subset of the whole dataset for IMDb. We report the average results of five runs and show the standard deviations in brackets.}
\label{tab:effectiveness_MPL_IMDb}
\scalebox{0.85}{
\begin{tabular}{|c|c|c|c|c|}
\hline
\textbf{\#Party}          & \textbf{$\epsilon$} & \textbf{Local party} & \textbf{\tfencrypted} & \textbf{\queqiao} \\ \hline
\multirow{3}{*}{3-party}  & 0.25                & $79.48\; (\pm 0.72)$ & $80.55\; (\pm 0.49)$                 & $80.20\; (\pm 0.51)$             \\ \cline{2-5} 
                          & 1                   & $83.84\; (\pm 0.32)$ & $84.46\; (\pm 0.67)$                 & $84.49\; (\pm 0.28)$             \\ \cline{2-5} 
                          & 2                   & $84.08\; (\pm 0.19)$ & $84.70\; (\pm 0.44)$                 & $84.35\; (\pm 0.17)$             \\ \hline
\multirow{3}{*}{5-party}  & 0.25                & $77.2\; (\pm 1.01)$  & -                                    & $77.46\; (\pm 0.82)$             \\ \cline{2-5} 
                          & 1                   & $80.33\; (\pm 0.26)$ & -                                    & $84.01\; (\pm 0.15)$             \\ \cline{2-5} 
                          & 2                   & $80.69\; (\pm 0.21)$ & -                                    & $84.59\; (\pm 0.37)$             \\ \hline
\multirow{3}{*}{10-party} & 0.25                & $72.00\; (\pm 1.81)$ & -                                    & $75.48\; (\pm 1.03)$             \\ \cline{2-5} 
                          & 1                   & $77.84\; (\pm 1.30)$ & -                                    & $84.64\; (\pm 0.13)$             \\ \cline{2-5} 
                          & 2                   & $78.22\; (\pm 0.69)$ & -                                    & $84.56\; (\pm 0.31)$             \\ \hline
\end{tabular}}
\end{table}

\vspace{-2mm}
\section{Supplemental experimental results of Section~\ref{subsec:effectiveness_MPL}}~\label{missing_res_MPL}
In this section, we show the accuracy comparison results of global models and local models  for MNIST and IMDb. The results are shown in Table~\ref{tab:effectiveness_MPL_MNIST} and Table~\ref{tab:effectiveness_MPL_IMDb}. Like the results of CIFAR-10, the accuracy of the global models is higher than local models in all scenarios. Meanwhile, the accuracy gaps between global models and local models also become larger as the number of parties increases. These results further verify the effectiveness of training models on the logically global datasets that are composed of local datasets through MPL.

\section{\wqruan{Comparison of Feature Extraction  Methods } }~\label{acc_Feature}
\wqruan{In this section, we keep other experimental settings unchanged and compare our feature extraction method with the method proposed by Tramèr and Boneh~\cite{tramer2021differentially}. We run their source codes\footnote{\url{https://github.com/ftramer/Handcrafted-DP}} to extract the features of MNIST and CIFAR-10 datasets with methods described in their paper.}

\wqruan{We show the experimental results in Table~\ref{tab:acc_Feature}. Compared with differentially private models trained on the features extracted by Tramèr and Boneh method, the models trained on the features extracted by our method can achieve higher accuracy. Because the features extracted by Tramèr and Boneh method are high-dimension (3969 vs. 1800 (PEA) in MNIST, 4096 vs. 2048 (PEA) in CIFAR-10) and sparse, the information of these features that are represented as fixed-point number with limited precision might be erased when we perform truncation and gradient clipping with SMPC protocols. Specifically, the low accuracy of models trained by \tfencrypted is from the truncation error of \texttt{ABY3} protocol~\cite{mohassel2018aby3}, i.e. it might truncate a fixed-point number with a very large error. When the dimension of features becomes higher, the truncation error occurs with higher probability, which is also shown in one previous study~\cite{279898}.}

\begin{table}[ht]
\centering
\caption{\wqruan{Accuracy (in \%) comparison of  the models trained on the features extracted by our method (PEA) and the models trained on features extracted by Tramèr and Boneh method. We report the average results of five runs and show the standard deviations in brackets.}}
\label{tab:acc_Feature}
\scalebox{0.85}{
\begin{tabular}{|l|cc|cc|}
\hline
\multirow{2}{*}{}                                 & \multicolumn{2}{c|}{PEA}                                     & \multicolumn{2}{c|}{Tramèr and Boneh~\cite{tramer2021differentially}}                        \\ \cline{2-5} 
                                                  & \multicolumn{1}{c|}{MNIST}              & CIFAR-10           & \multicolumn{1}{c|}{MNIST}              & CIFAR-10           \\ \hline
\multicolumn{1}{|c|}{\textbf{\tfencrypted}} & \multicolumn{1}{c|}{95.14 ($\pm 0.09$)} & 88.99 ($\pm 0.08$) & \multicolumn{1}{c|}{9.80 ($\pm 0.00$)}  & 8.35 ($\pm 2.06$)  \\ \hline
\multicolumn{1}{|c|}{\textbf{\queqiao}}     & \multicolumn{1}{c|}{94.90 ($\pm 0.14$)} & 88.18 ($\pm 0.33$) & \multicolumn{1}{c|}{92.81 ($\pm 0.06$)} & 86.04 ($\pm 0.15$) \\ \hline
\end{tabular}
}
\end{table}

\end{document}